%% file: main.tex
\documentclass[sigconf, nonacm]{acmart}

\usepackage{pifont}
\usepackage{comment}
\usepackage{float}
\usepackage[utf8]{inputenc}
\usepackage{balance}
\usepackage{algpseudocode}
\usepackage{subcaption}
\usepackage{tabularx}
\usepackage{diagbox}
\usepackage{hyperref}
\usepackage[noabbrev, capitalise, nameinlink]{cleveref}
\usepackage{xcolor,colortbl}
\usepackage[vlined,noend,linesnumbered,ruled,resetcount]{algorithm2e}
\newcommand{\cmark}{\textcolor{green!80!black}{\ding{51}}}
\newcommand{\xmark}{\textcolor{red}{\ding{55}}}
\newcommand{\qmark}{\textcolor{black}{{\fontfamily{cyklop}\selectfont \textit{?}}}}
\input{math_commands}
\newcommand{\eat}[1]{}

\newcommand{\norm}[1]{\left\lVert#1\right\rVert}
\let\oldnl\nl
\newcommand{\nonl}{\renewcommand{\nl}{\let\nl\oldnl}}

\newtheorem{definition}{Definition}
\newtheorem{problem}{Problem}
\newtheorem{lemma}{Lemma}
\newtheorem{conjecture}{Conjecture}

\newtheorem{theorem}{Theorem}
\newtheorem{myexp}{Example}
\newtheorem{corollary}{Corollary}

\newcommand\vldbdoi{10.14778/3467861.3467870}
\newcommand\vldbpages{1805-1817}
\newcommand\vldbvolume{14}
\newcommand\vldbissue{10}
\newcommand\vldbyear{2021}
\newcommand\vldbauthors{\authors}
\newcommand\vldbtitle{\shorttitle} 
\newcommand\vldbavailabilityurl{https://github.com/yikai-wu/Multi-Analyst-DP}
\newcommand\vldbpagestyle{plain} 

\begin{document}

\title{Budget Sharing for Multi-Analyst Differential Privacy}

\author{David Pujol}
\affiliation{%
  \institution{Duke University}
  \city{Durham}
  \state{NC}
  \country{United States}
  \postcode{27708}
}
\email{dpujol@cs.duke.edu}

\author{Yikai Wu}
\affiliation{%
  \institution{Duke University}
  \city{Durham}
  \state{NC}
  \country{United States}
  \postcode{27708}
}
\email{yikai.wu@duke.edu}

\author{Brandon Fain}
\affiliation{%
  \institution{Duke University}
  \city{Durham}
  \state{NC}
  \country{United States}
  \postcode{27708}
}
\email{btfain@cs.duke.edu}

\author{Ashwin Machanavajjhala}
\affiliation{%
  \institution{Duke University}
  \city{Durham}
  \state{NC}
  \country{United States}
  \postcode{27708}
}
\email{ashwin@cs.duke.edu}

\begin{abstract}
\input{abstract.tex}
\end{abstract}

\maketitle

\pagestyle{\vldbpagestyle}
\begingroup\small\noindent\raggedright\textbf{PVLDB Reference Format:}\\
\vldbauthors. \vldbtitle. PVLDB, \vldbvolume(\vldbissue): \vldbpages, \vldbyear.\\
\href{https://doi.org/\vldbdoi}{doi:\vldbdoi}
\endgroup
\begingroup
\renewcommand\thefootnote{}\footnote{\noindent
This work is licensed under the Creative Commons BY-NC-ND 4.0 International License. Visit \url{https://creativecommons.org/licenses/by-nc-nd/4.0/} to view a copy of this license. For any use beyond those covered by this license, obtain permission by emailing \href{mailto:info@vldb.org}{info@vldb.org}. Copyright is held by the owner/author(s). Publication rights licensed to the VLDB Endowment. \\
\raggedright Proceedings of the VLDB Endowment, Vol. \vldbvolume, No. \vldbissue\ %
ISSN 2150-8097. \\
\href{https://doi.org/\vldbdoi}{doi:\vldbdoi} \\
}\addtocounter{footnote}{-1}\endgroup

\ifdefempty{\vldbavailabilityurl}{}{
\vspace{.3cm}
\begingroup\small\noindent\raggedright\textbf{PVLDB Artifact Availability:}\\
The source code, data, and/or other artifacts have been made available at \url{\vldbavailabilityurl}.
\endgroup
}

\balance
\section{Introduction}\label{sec:intro}
\input{introduction.tex}

\section{Background}\label{sec:background}
\input{background}

\section{Problem Formulation}
\label{sec:ProbDef}
\subsection{Setting}
\input{prob_def}

\subsection{Desiderata}
\label{sec:desiderata}
\input{desiderata.tex}

\subsection{Problem Statement}
\input{problem-statement.tex}

\section{Design Paradigms} \label{sec:Design}
\input{design_paradigm.tex}

\section{Adapting Existing Mechanisms} \label{sec:Algorithms}

\input{algorithms.tex}

\section{Experiments} \label{sec:Experiments}
\input{experiments.tex}

\section{Related Work}
\input{relatedwork.tex}

\section{Future Work}
\input{futurework.tex}

\section{Conclusion}
\input{conclusions.tex} \label{sec:Conclusion}

\begin{acks}
\input{acknowledgement.tex}
\end{acks}

\clearpage
\bibliographystyle{ACM-Reference-Format}
\bibliography{refs}

\end{document}

%% file: math_commands.tex
\usepackage{amsmath,amsfonts,amsthm,bm}

\def\eqref#1{equation~\ref{#1}}

\def\1{\bm{1}}

\def\vzero{{\bm{0}}}
\def\vone{{\bm{1}}}

\def\ve{{\bm{e}}}

\def\vu{{\bm{u}}}
\def\vv{{\bm{v}}}
\def\vw{{\bm{w}}}
\def\vx{{\bm{x}}}
\def\vy{{\bm{y}}}

\def\mA{{\bm{A}}}
\def\mB{{\bm{B}}}
\def\mC{{\bm{C}}}
\def\mD{{\bm{D}}}

\def\mI{{\bm{I}}}

\def\mM{{\bm{M}}}

\def\mP{{\bm{P}}}

\def\mS{{\bm{S}}}
\def\mT{{\bm{T}}}

\def\mW{{\bm{W}}}

\DeclareMathAlphabet{\mathsfit}{\encodingdefault}{\sfdefault}{m}{sl}
\SetMathAlphabet{\mathsfit}{bold}{\encodingdefault}{\sfdefault}{bx}{n}

\def\gA{{\mathcal{A}}}
\def\gB{{\mathcal{B}}}

\def\gH{{\mathcal{H}}}

\def\gM{{\mathcal{M}}}

\def\gW{{\mathcal{W}}}

\DeclareMathOperator{\Err}{Err}
\DeclareMathOperator{\Error}{Error}
\DeclareMathOperator{\Lap}{Lap}

\DeclareMathOperator{\Mat}{Mat}

\DeclareMathOperator{\MM}{MM}
\DeclareMathOperator{\cnorm}{cnorm}
\DeclareMathOperator{\Rows}{Rows}

\DeclareMathOperator{\similarity}{sim}

\SetKw{kwnew}{new}
\SetKw{kwexists}{exists}
\SetKw{kwst}{s.t.}

%% file: abstract.tex
Large organizations that collect data about populations (like the US Census Bureau) release summary statistics that are used by multiple stakeholders for resource allocation and policy making problems. These organizations are also legally required to protect the privacy of individuals from whom they collect data. Differential Privacy (DP) provides a solution to release useful summary data while preserving privacy. Most DP mechanisms are designed to answer a single set of queries. In reality, there are often multiple stakeholders that use a given data release and have overlapping but not-identical queries. This introduces a novel joint optimization problem in DP where the privacy budget must be shared among different analysts. 

We initiate study into the problem of DP query answering across multiple analysts. To capture the competing goals and priorities of multiple analysts, we formulate three desiderata that any mechanism should satisfy in this setting -- The Sharing Incentive, Non-Interference, and Adaptivity -- while still optimizing for overall error. We demonstrate how existing DP query answering mechanisms in the multi-analyst settings fail to satisfy at least one of the desiderata. We present novel DP algorithms that provably satisfy all our desiderata and empirically show that they incur low error on realistic tasks.

%% file: introduction.tex
Large data collecting organizations like Facebook, Google, The U.S. Census Bureau, and Medicare often release summary statistics about individuals and populations. Access to such data is incredibly useful for multiple resource allocation, policy-making and scientific endeavors. Decisions like congressional seat apportionment, school funding and emergency response plans all depend on census data \cite{Census_uses}. Facebook's trove of user interaction data was found to be valuable in studying the impact of social media on elections and democracy \cite{FacebookSocialScienceOneData}. 

While these data releases are very useful, they may reveal sensitive information about individuals \cite{haney2017,machanavajjhala_privacy_on_the_map,vaidya_2013}. Differential Privacy (DP) \cite{dwork2014:textbook,DiffPriv} is the gold standard of privacy protection through the addition of randomized noise. However, due to the fundamental law of information recovery \cite{dinur_nissim_2003},  making an unbounded number of releases from a dataset (even if each satisfies DP) will eventually allow an attacker to accurately reconstruct the underlying dataset. Because of this, data curators must bound the amount of information released using a parameter known as the privacy loss budget $\epsilon$. Traditional privacy mechanisms focus on minimizing the error introduced by differential privacy, where error trades off with $\epsilon$. \par 

\subsection{Multi-analyst DP data release problem} We study the common real-world situation where multiple stakeholders or analysts are interested in a particular data release and the data curator must decide how the stakeholders should share the limited privacy budget. Consider the role of Facebook in its partnership with Social Science One \cite{FacebookSocialScienceOne}. Facebook wanted to aid research on the effect of social media on democracy and elections by sharing some social network data. In order to participate and receive the privacy protected data  each analyst had to submit their specific tasks and queries ahead of time. With the given set of queries from each analyst and a fixed privacy budget, Facebook created a single data release to be used by all analysts. Using existing DP techniques, Facebook had three options: (a) split the privacy budget and answer each analyst's queries individually, (b) join all analysts' queries together and answer them all at once using a workload answering mechanism \cite{Xu2013,Matrix10,HDMM,AHP,Chen13:recursive,Ding2011,Hay2010,Li2014,Narayan,Hb,Acs2012,torkzadehmahani2020dpcgan}, or (c) generate a single set of synthetic data \cite{PrivBays} for all analysts to use.\par 

Option (a) is inefficient as the same query can be answered multiple times, each time using some of the privacy budget. Option (b) may be efficient with respect to overall error but does not differentiate between the queries of different analysts. Some analysts may receive drastically more error than others, perhaps much more than they would have under (a). Option (b) therefore lacks much in the way of guarantees to an individual analyst. Option (c) is agnostic to any analyst's particular queries and may incur inefficiencies due to its inability to adapt to the specific queries being asked.\par 

Though all of these techniques have their uses, they all have some undesirable properties in the multi-analyst setting. This is because almost all of the work in differential privacy up until now has focused (often implicitly) on the single analyst case.  We are interested in designing effective shared systems for multi-analyst differentially private data release that simultaneously provide guarantees to individual analysts and ensure good overall performance. We call this the multi-analyst differentially private data release problem.

\subsection{Contributions} Our work introduces the multi-analyst differentially private data release problem.  In this context we ask: \textit{``How should one design a privacy mechanism when multiple analysts may be in competition over the limited privacy budget''}. Our main contributions in this work are as follows. \par

\begin{itemize}
    \item We study (for the first time)  differentially private query answering across multiple analysts. We consider a realistic setting where multiple analysts pose query workloads and the data owner makes a single private release to answer all analyst queries. 
    \item We define three minimum desiderata that that we will argue any differentially private mechanism should satisfy in a multi-agent setting -- The Sharing Incentive, Non-Interference and Adaptivity.
    \item We show empirically that existing mechanisms for answering large sets of queries either violate at least one of the desiderata described or are inefficient.
    \item We introduce mechanisms which provably satisfy all of the desiderata while maintaining efficiency. 
\end{itemize}

%% file: background.tex
\noindent\textbf{Data Representation}
We consider databases where each individual corresponds to exactly one tuple. The algorithms considered use a vector representation of the database denoted $\vx$. More specifically, given a set of predicates $ \mathcal{B} = \{\phi_1 \dots \phi_n\}$, the original database $D$ is transformed into a vector of counts $\vx^D $ where $\vx_i^D$ is the number of records in $D$ which satisfy $\phi_i$. For simplicity, we denote the length of the data vector as $n$ and we will use the notation $\vx$ in order to refer to the vector form of database $D$  \par
\noindent\textbf{Predicate counting queries}
 are a versatile class of queries that count the number of tuples satisfying a logical predicate. A predicate corresponds to a condition in the \textbf{WHERE} clause of an SQL query. So a predicate counting query is one of the form \textbf{SELECT Count (*) FROM R WHERE $\phi$}. Workloads of counting queries can express queries such as histograms, high dimensional range queries, marginals, and datacubes among others.  \par 
Like databases, a predicate counting query can be represented as a $n$-length vector $\vw$ such that the answer to the query is $\vw^T\vx$.
A workload is a set of $m$ predicate counting queries arranged in a $m \times n$ matrix $\mW$, where each row is the vector form of a single query. Many common queries can be represented as workloads in this form. For example, a histogram query is simply represented by an $n \times n$ identity matrix.

\noindent\textbf{Differential Privacy}
 \cite{dwork2014:textbook,DiffPriv} is a formal model of privacy that grantees each individual that any query computed from sensitive data would have been almost as likely as if the individual had opted out. More formally, Differential Privacy is a property of a randomized algorithm which bounds the ratio of output probabilities induced by changes in a single record. 
\begin{definition}[Differential Privacy]
A rand\-om\-ized mechanism $\gM$ is $(\epsilon$,$\delta$ )-differentially private if for two neighboring databases $D$, and $D'$ which differ in at most one row, and any outputs $O \subseteq Range(\gM)$:
\begin{equation*}
\Pr[ \gM(D) \in O] \leq \exp(\epsilon) \times \Pr[ \gM(D') \in O] + \delta
\end{equation*}
\end{definition}
 
The parameter $\epsilon$ often called the privacy budget quantifies the privacy loss. Here we focus exclusively on $\epsilon$-Differential Privacy, i.e when $\delta = 0$. \par The Laplace Mechanism is a differentially private primitive which underlines the algorithms used here. We describe the vector version of the Laplace Mechanism below. 
\begin{definition}[Laplace Mechanism, Vector Form]
Given an $m \times N$ workload matrix $\mW$, the randomized algorithm which outputs the following vector is $\epsilon$-differentially private \cite{dwork2014:textbook}. 
\begin{equation*}
\mW\vx + \Lap\left(\frac{\|\mW\|_1}{\epsilon}\right)^m
\end{equation*}
\end{definition}
Where $\| \mW\|_1$ is the maximum L1 column norm of $\mW$ and $\Lap(\sigma)^m$ denotes the $m$-length vector of $m$ independent samples from a Laplace distribution with mean $0$ and scale $\sigma$.

Differentially private releases compose with each-other in that if there are two private releases of the same data with two different privacy budgets the amount of privacy lost is equivalent to the sum of their privacy budgets. More formally we have the following. 
\begin{theorem}[DP composition \cite{dwork2014:textbook}]
\label{thrm:composoition}
Let $\gM_1$ be an $\epsilon _1$-differentially private algorithm and $\gM_2$ be an $\epsilon _2$-differentially private algorithm. Then their combination defined to be $\gM_{1,2}(x) = (\gM_1(x), \gM_2(x))$ is $\epsilon _1 + \epsilon _2$-differentially private
\end{theorem}

Of the many algorithms proposed in the literature, we will consider a class of measures that invoke the \textbf{Select, Measure, Reconstruct} paradigm, where instead of directly answering the queries, they first \textbf{select} a new set of strategy queries. They then \textbf{measure} the strategy queries using a privacy protecting mechanism (in this case the Laplace Mechanism \cite{dwork2014:textbook}) and finally \textbf{reconstruct} the answers to the original input queries from the noisy measurements. Examples of mechanisms that follow this paradigm are the Matrix Mechanism \cite{Matrix} and it's derivatives such as HDMM\cite{HDMM}. The Matrix Mechanism answers a workload of queries $\mW$ by first selecting a strategy workload $\mA$ to answer. It then measures the queries in $\mA$ using the Laplace Mechanism and then reconstructs the answers to $\mW$ from the noisy answers of $\mA$. Given a workload matrix $\mW$ and a strategy matrix $\mA$, the expected total square error of the Matrix mechanism is as follows. 
\begin{equation} \label{eqn:error}
\Error(\mW,\mA, \epsilon) = \frac{2}{\epsilon ^2} \| \mA\|_1^2 \| \mW\mA^+ \|^2_F    
\end{equation}
Where $\| \mA\|_1$ is the L1 column norm of $\mA$ and the norm considered here is the frobenius norm. For concreteness in this work we consider only mechanisms which answer workloads of linear queries. These mechanisms can be extended to answer non-linear queries by adding post-processing steps which reconstruct non-linear queries from answers to several linear queries.

%% file: prob_def.tex
 We consider the setting where there are $k$ analysts with associated positive weights $s_1, s_2 \dots s_k \in (0,1)$ such that $ s_1 + s_2 \dots s_k = 1$. These weights represent the share of the total privacy budget to which each analyst is entitled and can be interpreted as the relative importance of each analysts' queries; the natural default is to use proportional weights of $1/k$ for every analyst. 
\par 
Each analyst submits a workload of queries $W_1, W_2 \dots W_k \in \mathcal{W}$  The data curator then answers all of the queries using a multi-analyst differentially private mechanism. We define a  multi analyst differentially private mechanism $\mathcal{M}$ as a function that takes as input each analysts' set of queries, their respective shares of the privacy budget and the overall privacy budget and outputs a single data release containing the answers to all of the queries.

We can describe the mean squared error experienced by a particular analyst in a multi-analyst Matrix Mechanism as follows.
\begin{equation}
    \Err_i(\mathcal{M},\mathcal{W},\epsilon) = \frac{2}{\epsilon ^2} \| \mA\|_1^2 \| \mW_i\mathbf{A^+} \|^2_F,    
\end{equation}
where $\mW_i$ is the matrix form of the workload $W_i$ given by the $i$th analyst, $\mA$ is the strategy matrix produced by mechanism $\mathcal{M}$ with input $\mathcal{W}$. This formula is only for linear queries. For non-linear queries, we must use real datasets to get query answers and estimate expected errors.
\par

%% file: desiderata.tex
For ease of exposition, imagine that each analyst is given the choice to either have their queries answered independently with their share of the privacy budget or to join the collective, a group of analysts whose queries are answered with a multi-analyst DP mechanism using the sum of all of the collective analysts' privacy budget. We argue that any multi-analyst differentially private mechanism should satisfy three desiderata. First, the mechanism should incentivize a rational agent to participate in the collective by guaranteeing no worse expected error than if their queries were answered independently. Second, the mechanism should never cause any analyst to regret that another analyst is participating in the collective and increasing the former's expected error. Third, the mechanism should be able to adapt to and optimize for the particular queries being asked by all analysts. In this section we formalize these criteria through three separate desiderata: the Sharing Incentive, Non-Interference, and Adaptivity. We introduce each of the desiderata as well as current common practice through a rolling example which demonstrates the importance of these desiderata even in a simple case. \par 

 \begin{myexp} \label{Example:setup}
 Alice, Bob, and Carol are analysts working on a private dataset of US COVID-19 deaths by age provided by the Center for Disease Control \cite{CDC_Covid_counts}. The populations are split into $11$ buckets by age. The data curator decides to use a privacy budget of $ \epsilon = 1$. Each of the analysts are entitled to an equal share of the privacy budget (that is, each has weight $1/3$). Alice and Bob both  want to ask the a histogram of the counts by age ( we call this the identity workload on age). Carol wants to ask for the total of all counts in the database. 
 \end{myexp}

 The first desideratum, the \textbf{Sharing Incentive} requires that each analyst, in expectation, receives at most as much error as if they had computed their query answers independently using the same mechanism and their fraction of the privacy budget. This captures the idea that each analyst should always benefit from joining the collective.
\begin{definition}[Sharing incentive] \label{SharingIncentive}
A mechanism $\gM$ satisfies the Sharing Incentive if for every analyst $i$ the following holds. 
\begin{equation*}
\Err_i(\gM,\gW,\epsilon) \leq  \Err_i(\gM,\{W_i\},s_i \epsilon)
\end{equation*}
\end{definition}

\begin{myexp} \label{Example:sharing}
The data curator decides to split each analyst off and give them each $\epsilon/3$ of the privacy budget in order to answer their queries independently using HDMM. In this case  Alice and Bob both receive a total expected error of $\pm 198 $ people while Carol receives an expected error of $\pm 18$ people.

Suppose the data curator decides to pool the queries and jointly answer them using HDMM. Alice and Bob receive $\pm 22$  as expected their error which is less than their error using the independent mechanism. Carol received $\pm 22$  as her expected error which is more error than in the independent case where her expected error was $\pm 18$ thus violating the Sharing Incentive.  \par 
\end{myexp}

In this case Carol would prefer her workload to be answered independently while Alice and Bob would join together. If the mechanism were to satisfy the Sharing Incentive, Carol would be guaranteed no worse error by joining Alice and Bob and as such should always make that choice.  \par 

The second desiderata is \textbf{Non-Interference}, which states that adding an additional analyst to the collective group, with their associated share of the privacy budget, should not increase the error experienced by any of the analysts already in the collective. This desiderata ensures that no analyst in the collective can ask (intentionally or unintentionally) a malicious set of queries which would increase the error of any of the other analysts more than if they had never joined the collective. Likewise, Non-interference ensures that adding more analysts to the collective (and with them more privacy budget) can only improve the accuracy of all agents.
\begin{definition}[Non-interference]\label{def:non-int}
A mechanism $\gM$ satisfies Non-Interference if for all analysts $i \neq j$, for all workloads $W_i,W_j$ \\ \begin{equation*}
    \Err_i(\gM, \gW, \epsilon) \leq \Err_i(\gM, \gW \setminus W_j, (1-s_j)\epsilon)
\end{equation*} 
\end{definition}
\begin{myexp} \label{example:interference}
Alice and Bob have decided to join the collective and answer their queries together since they have the same queries. They run the joint mechanism on their queries using $\frac{2 }{3}\epsilon$ of the budget. Here they both receive $\pm 22$ people as expected error. Carol then joins the collective. They then rerun the same mechanism using the entire budget. In this case, Alice and Bob receive an expected error of $\pm 24 $ people,  which is more than their original $\pm 22$ people expected error therefore violating Non-Interference.
\end{myexp}

In this case, Carol joining the collective makes both Alice and Bob worse off. If the mechanism were to satisfy Non-Interference, Alice and Bob would be guaranteed that no matter what workload Carol asks they can be to be no worse off for allowing Carol into the collective. 

\par

Our third desideratum is \textbf{Adaptivity}, which states that a mechanism should be able to adapt to the inputs given. We say that a mechanism is adaptive if it changes its query answering strategy based off all the inputs given. This ensures that a mechanism can adapt to the specific queries being asked by analysts in order to avoid high error for particular sets of queries.

\begin{myexp} \label{Example:adaptivity}
The data curator chooses to use a non-adaptive mechanism which always releases data by answering the Identity workload. Alice and Bob are happy since this is their exact workload. Carol is punished since her query workload cannot be efficiently reconstructed using the identity workload and receives an expected error of $\pm 24 $ people,  which is worse than her independent expected error of $\pm 18$ people.
\end{myexp}
An adaptive mechanism would be able to adapt it's query answering strategy in order to account for Carol's queries therefore reducing her error. The concept of Adaptivity highlights the flaws of various trivial mechanisms which satisfy the Sharing Incentive and Non-Interference by intentionally ignoring the inputs or interactions between analysts inputs.\par

\noindent\textbf{Tradeoffs Between Desiderata and Accuracy.}
\label{sec: Tradeoffs}
Both the Sharing Incentive and Non-interference add additional constraints to mechanisms in the multi-analyst setting. As such, we expect that mechanisms which satisfy these desiderata will suffer some accuracy loss. In contrast, adaptivity is not in conflict with accuracy. Rather adaptivity is a requirement that a  mechanism should optimize it's query strategy to be more efficient for a given workload. Overall, we expect a mechanism that is adaptive to perform better over a wide range of queries as opposed to its non-adaptive counterpart. \par 

%% file: problem-statement.tex
Our goal is to design multi-agent differentially private mechanisms that answer the workloads submitted by the analysts with low error while satisfying the three desiderata -- sharing incentive, non-interference and workload adaptivity. More formally: 

\begin{problem}
Given any $k$ workloads $W_1, \ldots, W_k$ of queries on a database $D$ with weights $s_1, \ldots, s_k \in (0,1)$ s.t. $s_1+ \ldots + s_k = 1$, design an adaptive mechanism $\mathcal{M}$ such that: 
\begin{itemize}
    \item $\mathcal{M}$ satisfies $\epsilon$-differential privacy, and
    \item  $\mathcal{M}$ satisfies sharing incentive (\cref{SharingIncentive}), non-interference (\cref{def:non-int}) .
\end{itemize}
\end{problem}

%% file: design_paradigm.tex
\begin{figure*}[ht]
\resizebox{0.75\textwidth}{!}{%
	\centering
	\begin{subfigure}[b]{0.30\textwidth}
		\includegraphics[width=1\textwidth]{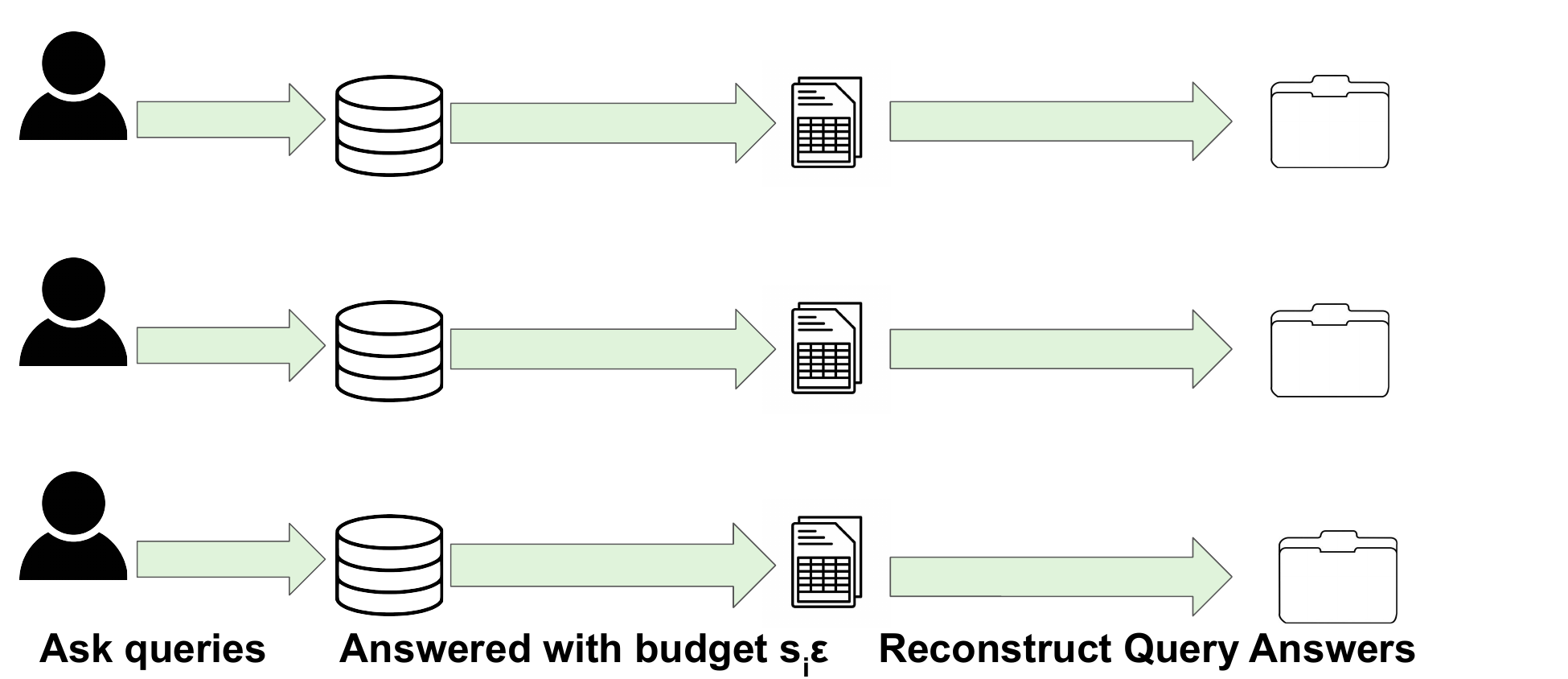}
		
		\caption{Independent}\label{fig:independent}
	\end{subfigure}
	\begin{subfigure}[b]{0.30\textwidth}
		\includegraphics[width=1\textwidth]{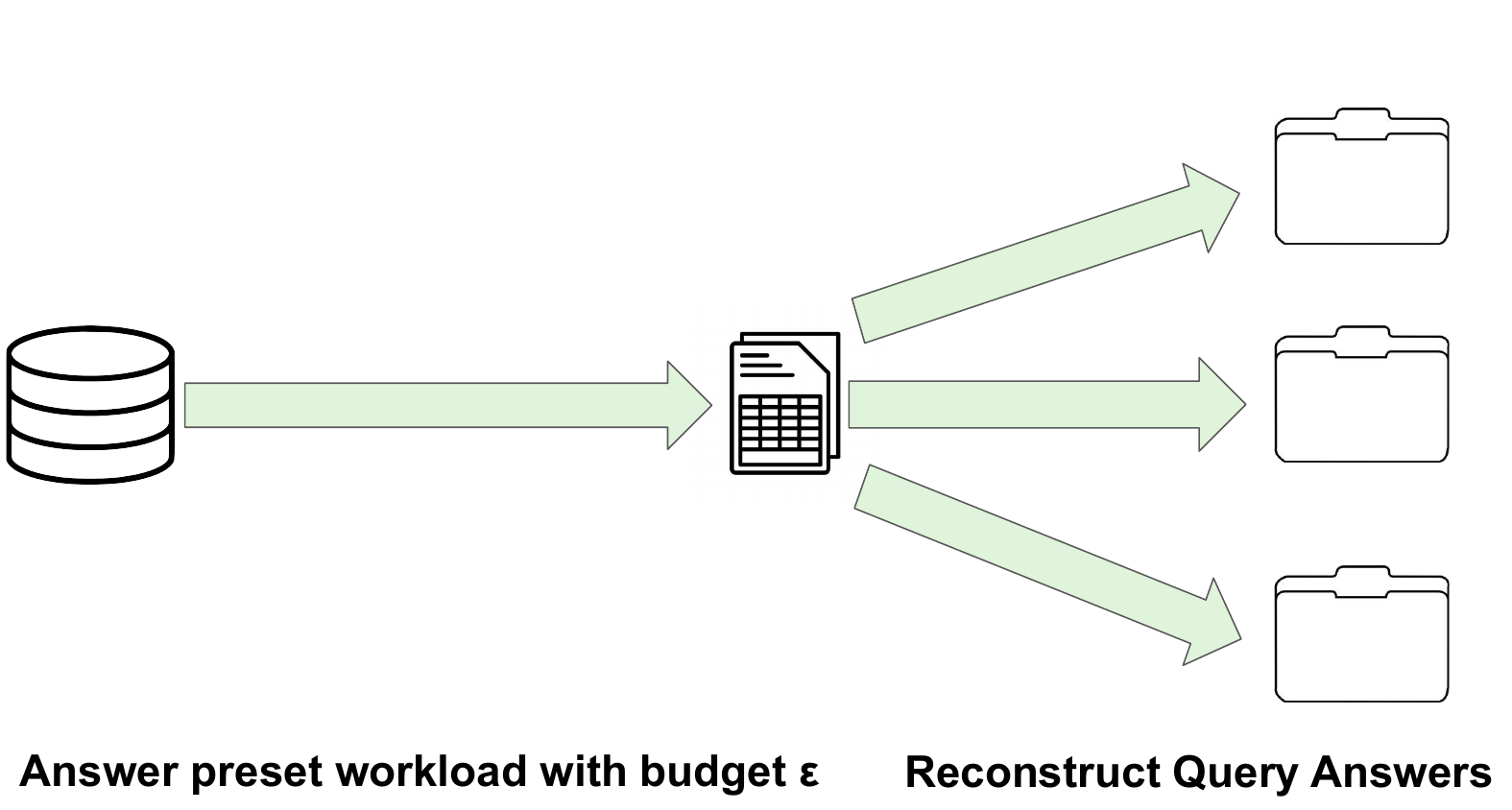}
		
		\caption{Workload Agnostic}\label{fig:agnostic}
	\end{subfigure}
}
\resizebox{0.75\textwidth}{!}{%
    \begin{subfigure}[b]{0.30\textwidth}
		\includegraphics[width=1\textwidth]{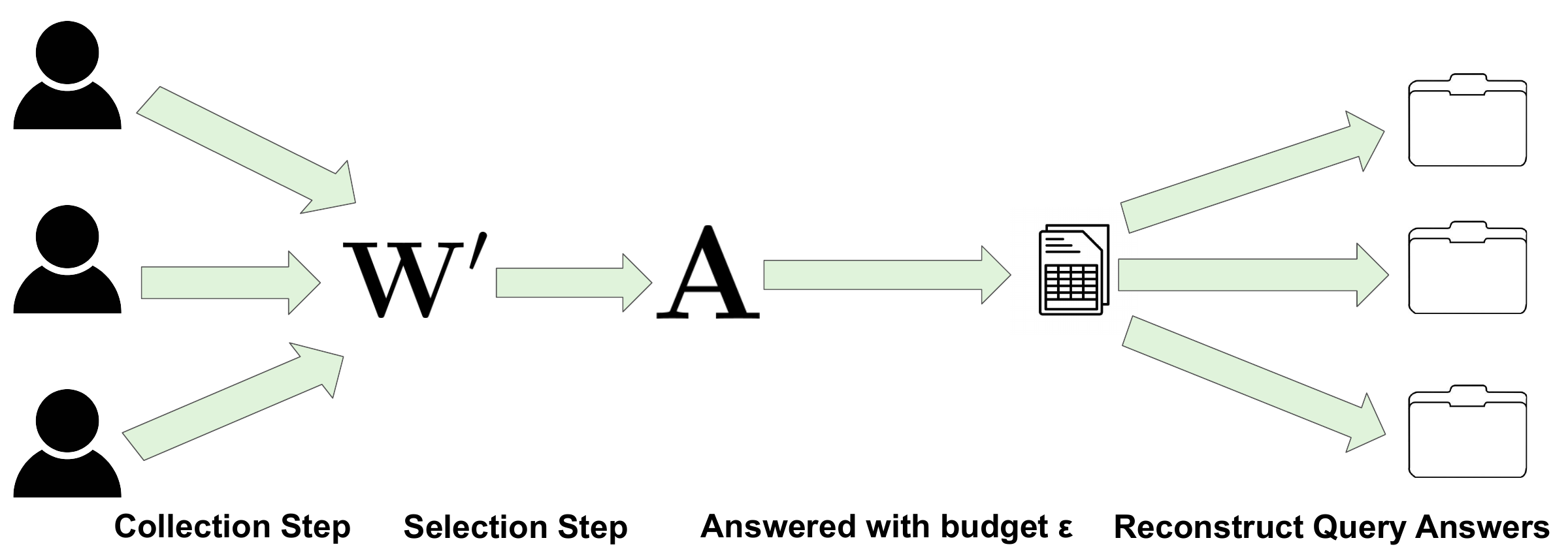}
		
		\caption{Collect First}\label{fig:collect}
	\end{subfigure}
	\begin{subfigure}[b]{0.30\textwidth}
		\includegraphics[width=1\textwidth]{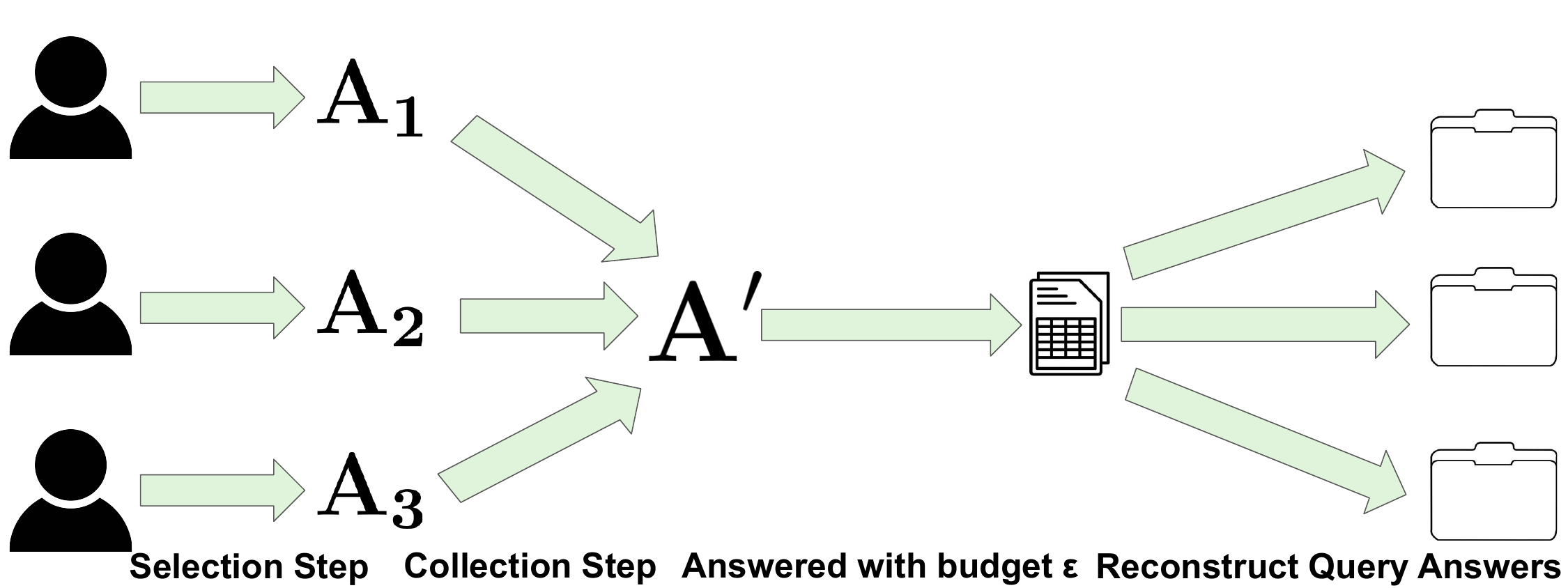}
		
		\caption{Select First}\label{fig:select}
	\end{subfigure}
}
\caption{Design Paradigms for Multi-Analyst DP Query Answering}
\label{fig:algos}
\end{figure*}

Here we introduce 4 design paradigms which we use to guide our design of multi analyst differentially private mechanisms. We call the first two classes Independent and Workload Agnostic. These classes use existing mechanisms without explicitly considering the group structure of the problem. We consider them as baselines for comparison; it is easy to see in theory and we show empirically that these mechanisms lead to poor performance with respect to total error. We call the other two classes of mechanisms Collect First and Select First. These mechanisms adapt the Select Measure Reconstruct Paradigm by aggregating all analysts' queries either before or after the selection step respectively. Each of the paradigms are depicted in \cref{fig:algos}. 

\noindent\textbf{Independent Mechanisms} give each analyst their share of the overall privacy budget proportional to their weights $s_1, \hdots s_k$ and answers each analyst's queries independently of one another using some workload answering mechanism. Mechanisms of this class by definition satisfy both the Sharing Incentive and Non-Interference since analysts always have the same expected error regardless of how many analysts are in the collective or what their queries are.

\begin{lemma}
Any Independent Mechanism satisfies both the Sharing Incentive and Non-Interference
\end{lemma}
These mechanisms are not efficient as they typically answer each individual query with less privacy budget than other mechanisms and may answer the same or similar queries multiple times. In \cref{sec:waterfilling} we will show a mechanism that satisfies all the desiderata and can achieve up to $\sqrt{k}$ times better error than its independent counterpart. 

\noindent\textbf{Workload Agnostic Mechanisms} always answer the same set of queries with the entire budget regardless of the analysts' workloads. Mechanisms of this class also trivially satisfy both the Sharing Incentive and Non-Interference since the same workload is always answered regardless of the preferences of the analysts. Joining the collective only increases the overall privacy budget leading to an overall decrease in error, satisfying the sharing incentive. Likewise, whenever a new analyst joins the collective the workload remains the same and the privacy budget increases causing an overall decrease in error for all analysts, therefore satisfying Non-Interference.

\begin{lemma}
Any Workload Agnostic mechanism satisfies both the Sharing Incentive and Non-Interference
\end{lemma}
  Workload Agnostic Mechanisms are not adaptive and this causes them to be inefficient with respect to total error, even for a single analyst. For example, if a noisy count was released for people of ages $\{ 0,1, 2 \dots 99\}$ but an analyst asks for the total count of all people then the answer to the total query, reconstructed by adding together all of the noisy counts, has at least 10 times larger error than if the total query was answered directly using all of the privacy budget. \par 

\noindent\textbf{Collect First Mechanisms}  collect all analysts' queries together before the selection step. These mechanisms combine all of the workloads of each analyst into some weighted query set, then run the selection step to select a single strategy workload for all the analysts' workloads.\par 

\noindent\textbf{Select First Mechanisms} collect all the analysts' queries after the selection step in a Select Measure Reconstruct mechanism. Mechanisms of this class allow an individual strategy for each analyst's queries. After a strategy is selected for each analyst's queries they are all aggregated into a joint strategy workload which is answered directly.

The fundamental problem with designing a Collect-First Mechanism that satisfies the sharing incentive and Non-Interference is that it is difficult to reason about or enforce the properties on any useful selection step (such as HDMM) optimizing on the joint workload. Select-First Mechanisms are  easier to work with because we do not need to reason over the multi-analyst properties of the selection step when it is applied to each analyst independently first.\par

%% file: algorithms.tex
\begin{table}[ht]
\caption{Desiderata satisfied by algorithms}
\begin{tabularx}{\columnwidth}{|l|X|X|X|}
    \hline
     \diagbox[width=0.36\hsize]{\textbf{Mechanism}}{\textbf{Desiderata}} &  Sharing\newline Incentive  & Non-interference & Adaptivity\\
    \hline
    Independent  & \cmark   & \cmark & \cmark  \\
    \hline
    Identity  & \cmark & \cmark & \xmark\\
    \hline
    Utilitarian & \xmark & \xmark & \cmark\\
    \hline
    Weighted Utilitarian  & \qmark  & \xmark & \cmark \\
    \hline
    0-Waterfilling  & \cmark & \cmark & \cmark \\
    \hline
\end{tabularx}
\label{tab:algos}
\end{table}

We introduce and analyze four mechanisms for multi-analyst query answering. Each of these mechanisms directly invokes one of the design paradigms listed in \cref{sec:Design}. Of these mechanisms, the Independent Mechanism and Identity Mechanism are both direct applications of existing mechanisms to the multi-analyst setting. The Utilitarian Mechanism and Weighted Utilitarian Mechanism are adaptations of HDMM\cite{HDMM} to the multi-analyst setting. The desiderata satisfied by algorithms are shown in \cref{tab:algos}

All of the properties and proofs given hold for linear queries which are answered directly by the mechanisms. These properties do not inherently hold for any non-linear queries reconstructed from linear query answers. In \cref{sec:experiments_nonlinear_results} we empirically study our mechanisms answering non-linear queries such as median and percentiles.\par 

\noindent\textbf{Independent HDMM} invokes the Independent Mechanism paradigm and simply runs HDMM\cite{HDMM} for each analyst using their share of the privacy budget. We use this mechanism as a baseline to compare other mechanisms to in  \cref{sec:Experiments}. As an example of an Independent Mechanism, it satisfies all three of the desiderata.

\noindent\textbf{Identity Mechanism}, as shown in algorithm \ref{alg:ind}  answers the identity strategy regardless of the preferences of the individual analysts. As an example of a workload agnostic mechanism it satisfies the Sharing Incentive and Non-Interference but is non-adaptive.\par

 \noindent\textbf{The Utilitarian Mechanism} is an example of a collect first mechanism. The Utilitarian Mechanism first aggregates each analysts' queries by creating a multi-set of queries which contains all the analysts' queries with multiplicity equal to the number of analysts asking the query. We then run a selection step on this joint query set. In general, we would expect this mechanism to achieve the minimum expected total error across all analysts, but the mechanism can easily violate both the Sharing Incentive and Non-Interference.
 
 \begin{algorithm}[ht]
\SetAlgoLined
\SetKwInOut{Input}{input}\SetKwInOut{Output}{output}
\Input{Set of $k$ workloads  $\gW \leftarrow \{W_1, W_2, \ldots, W_k\ \} $,\\ 
Set of $k$ budget weights $S \leftarrow \{s_1, s_2, \dots, s_k\ \} $ , \\
Data vector $\vx$,\\
privacy budget $\epsilon$,\\
Selection Mechanism $\gM$} 

 \nonl \textbf{Selection Step} \\
 $\gA \leftarrow \{\gM(W_i) \;|\; W_i \in \gW\}$\\

\nonl \textbf{Measure step} \\
$Y \leftarrow \{\mA_i\vx + \Lap(\frac{1}{s_i\epsilon}\norm{\mA_i}_1) \;|\; \mA_i \in \gA, s_i \in S\}$\\

\nonl \textbf{Reconstruct step} \\
$\Bar{X} \leftarrow \{\mA_i^+ \vy_i\;|\; \mA_i \in \gA ,  \vy_i \in \vy \}$\\ \nonl \Comment{$\mA^+$ is the Moore–Penrose inverse of $\mA$ and $\mA^+\mA = \mI$.}\\
ans $\leftarrow \{W_i(\Bar{\vx}_i)\;|\; W_i \in \gW, \Bar{\vx}_i \in \Bar{X}\}$\\

\Return ans
 \caption{Independent Mechanism}
\label{alg:ind}
\end{algorithm}

\begin{algorithm}[ht]
\SetAlgoLined
\SetKwInOut{Input}{input}\SetKwInOut{Output}{output}
\Input{ $\gW,S, \vx,\epsilon,\gM$ \Comment{defined in Algorithm~\ref{alg:ind}}}

 \nonl \textbf{Selection Step} \\
 $\mA \leftarrow \text{Identity}(n)$\\

\nonl \textbf{Measure step} \\
$\vy \leftarrow \mA\vx + \Lap(\frac{1}{\epsilon})$\\

\nonl \textbf{Reconstruct step} \\
$\Bar{\vx} \leftarrow \mA^+ \vy$\\
ans $\leftarrow \{W_i(\Bar{\vx})\;|\; W_i \in \gW\}$\\

\Return ans
 \caption{Identity Mechanism}
 \label{alg:iden}
\end{algorithm}

 \begin{theorem}
 The Utilitarian Mechanism does not satisfy the sharing incentive
 \end{theorem}
 \begin{proof}
 Take the general case from \cref{Example:sharing} where we have $k$ analysts. $k-1$ of those analysts ask the Identity workload (a histogram of all counts). The last analyst asks the Total workload  (a sum of all counts). Each analyst receives an equal share $\frac{\epsilon}{k}$ of the privacy budget. In the independent case each analyst asking the Identity workload would receive identity as a strategy and would experience $\left(\frac{2k}{\epsilon}\right)^2n$ error. The one analyst asking total would receive the Total workload  as their strategy and experience $\left(\frac{2k}{\epsilon}\right)^2$ error. If all the analysts join the collective an optimal utilitarian mechanism would chose the Identity workload as the workload that optimizes on total error. In this case (now using the entire privacy budget) each analyst would receive $\left(\frac{2}{\epsilon}\right)^2n$ error. In this case all the analysts asking the identity workload would benefit while the analyst asking the Total workload  will get increasingly worse error as $n$ (the size of the database) increases and will violate sharing incentive when $ k^2 < n$.
 \end{proof}
 \begin{theorem}
 The Utilitarian Mechanism does not satisfy non-interference
 \end{theorem}
\begin{proof}
Consider the case where there are $k$ analysts each with an equal $\frac{\epsilon}{k}$ share of the privacy budget. $k-1$ of these analysts ask the Total workload  and the last analyst asks the identity workload. If the $k-1$ analysts asking the Total workload  are in the collective the strategy used would directly answer the Total workload  and receive $\left(\frac{2k}{(k-1)\epsilon}\right) ^2$ expected error. If the last analyst were to join an optimal utilitarian mechanism would answer the queries using the identity strategy which optimizes on overall error. This would result in the $k-1$ analysts each receiving $\left(\frac{2}{\epsilon}\right)^2n$ expected error which violates non interference when $ \left( \frac{k}{k-1}\right)^2 < n$
\end{proof}
 
 \noindent\textbf{The Weighted Utilitarian Mechanism} is a variant of the Utilitarian Mechanism that attempts to directly optimize for the Sharing Incentive. This is achieved by weighting the queries prior to the collection step. This requires an additional set of $k$ parameters which we call workload weights $\Omega = \{\omega_1, \omega_2, \ldots, \omega_k\}$, where $\omega_i$ is the weight for workload $W_i$.
 After weighting each of the queries, the Utilitarian Mechanism is run on the weighted query sets. The Utilitarian Mechanism is a special case of Weighted Utilitarian Mechanism where $\omega_1 =\omega_2 = \cdots = \omega_k =1$.

In an attempt to satisfy the Sharing Incentive we set the weights as the inverse of the expected error of the mechanism in the independent case.
\begin{equation}
    \omega_i = \Err_i(\gM, W_i, s_i\epsilon)^{-1}
\end{equation}
These weights incentivize an optimizer to satisfy the sharing incentive as an analyst's utility is above 1 only if they have less error than required to satisfy the sharing incentive. We see in \cref{sec:Experiments} that these weights allow for the utilitarian mechanism to satisfy the sharing incentive in practical settings and we have not been able to create settings where the sharing incentive is violated. It is unclear if it satisfies the Sharing Incentive in all settings. 
\begin{conjecture}
  The weighted utilitarian mechanism satisfies the sharing incentive
\end{conjecture}
In \cref{sec:prac_results} we are able to show empirically that the weighted utilitarian mechanism does violate non interference. 
\begin{theorem}
The weighted utilitarian mechanism does not satisfy non-interference
\end{theorem}
  \begin{algorithm}[ht]
\SetAlgoLined
\SetKwInOut{Input}{input}\SetKwInOut{Output}{output}
\Input{$\gW,S, \vx,\epsilon,\gM$, \Comment{defined in Algorithm~\ref{alg:ind}}\\ 
Set of $k$ workload weights $\Omega \leftarrow \{\omega_1, \omega_2, \dots, \omega_k\} $}

 \nonl \textbf{Collection Step} \\
  $\gW' \leftarrow  \bigcupplus_{i=1}^k\omega_iW_i$\Comment{$\bigcupplus$ is multi-set union}\\

   \nonl \textbf{Selection Step} \\
 $\mA \leftarrow \gM(\gW') $ \\

\nonl \textbf{Measure step} \\

$\vy \leftarrow \mA\vx + \Lap(\frac{1}{\epsilon}\norm{\mA}_1)$\\

\nonl \textbf{Reconstruct step} \\
$\Bar{\vx} \leftarrow \mA^+ \vy$\\
ans $\leftarrow \{W_i(\Bar{\vx})\;|\; W_i \in \gW\}$\\

\Return ans
\caption{Weighted Utilitarian Mechanism}
\end{algorithm} 

\section{The Waterfilling Mechanism}
\label{sec:waterfilling}

\begin{figure*}[t]
   \centering
    \includegraphics[width=0.7\textwidth]{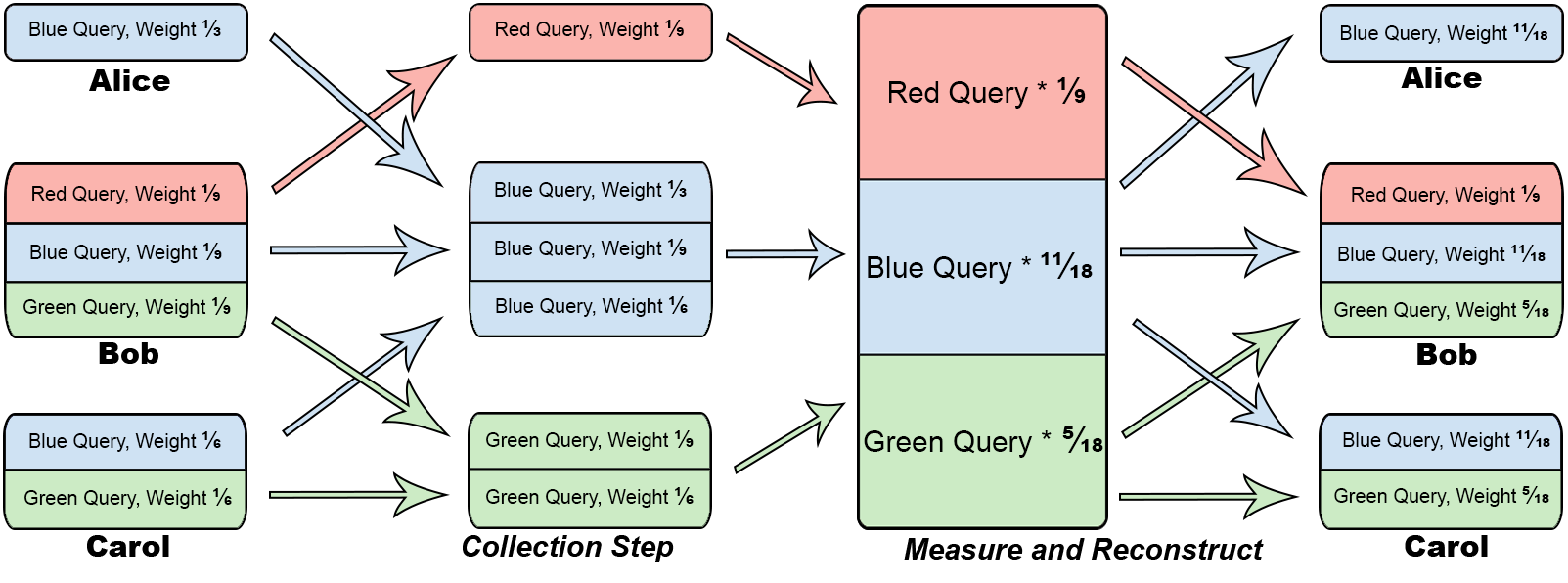}
    \caption{Simplified Waterfilling Mechanism}
    \label{fig:waterfill_animation}
  \end{figure*}

The Waterfilling Mechanism is an example of a select first mechanism which satisfies all  three of the desiderata. We first start with a simplified example of the Waterfilling Mechanism seen in \cref{fig:waterfill_animation} and then discuss the full Waterfilling Mechanism. \par 

In this example there are three analysts Alice, Bob, and Carol each given the same share of the budget, $\frac{1}{3}$. Alice asks only the blue query and assigns all of her share to that query. Bob asks the red, blue, and green queries and assigns each query equal amounts of his share of the privacy budget. Carol asks the blue and green queries and like Bob assigns his share of the budget equally across all her queries. The Waterfilling mechanism then buckets similar queries (in this example by bucketing red blue and green queries)  and their associated shares of privacy budget together.  Once all the queries are assigned to buckets the mechanism answers a single query for each bucket using the entire privacy budget in each bucket. The mechanism then uses those answered queries to reconstruct the analysts original queries. In \cref{fig:waterfill_animation}, we can see that since the red query was only asked by one analyst it receives the same amount of privacy budget as if were asked independently. Meanwhile since each analyst asked the blue query it is answered once using the pooled contribution of privacy budget from each analyst, resulting in a more accurate estimate than if each analyst had independently answered the blue query, even if they subsequently shared their results with one another. \par 

The example shown in \cref{fig:waterfill_animation} is a simplified version of the Waterfilling Mechanism. The Waterfilling Mechanism as defined in Algorithm \ref{alg:waterfilling} has three key differences. The first key difference is the selection step. In the simplified Waterfilling Mechanism analyst's queries are bucketed directly. However in practice a selection step is done first. This selection step takes in the analyst's workload and outputs a strategy workload that may be more efficient to answer directly. The second key difference is sensitivity scaling. The simplified example assumes that the sensitivity of each query is 1 and that all three queries overlap somewhat causing Alice's sensitivity to be 1 Bob's sensitivity to be 3 and Carols sensitivity to be 2. In order to avoid sensitivity scaling issues, the Waterfilling Mechanism scales each analyst's strategy workload to have a sensitivity of 1 prior to the bucketing step. The third key difference is in the bucketing step. In the simplified example we only bucketed identical queries. Since the selection step introduces some numerical instability we allow for queries which are approximately equal to be added to the same bucket. We introduce an additional parameter $\tau$ which determines how much two queries are allowed to deviate to be assigned to the same bucket. In Algorithm \ref{alg:waterfilling} we allow two queries with cosine similarity greater than $1-\tau$ to be assigned to the same bucket. Once the buckets are filled the query answered is the unit vector representing the average query in the bucket.

All of the proofs below assume that $\tau =0$ and may not hold for higher values of $\tau$. We set $\tau$ to be $10^{-3}$ in experiments and we empirically evaluate the performance of the Waterfilling mechanism as $\tau$ changes in \cref{sec:Experiments_Tolerance_results}.

\begin{algorithm}[ht]
\SetAlgoLined
\SetKwInOut{Input}{input}\SetKwInOut{Output}{output}
\Input{$\gW,S, \vx,\epsilon,\gM$, \Comment{defined in Algorithm~\ref{alg:ind}}\\
tolerance parameter $\tau$}

 \nonl \textbf{Selection Step} \\
 $\gA \leftarrow \{\gM(W_i) \;|\; W_i \in \gW\}$\\

 \nonl \textbf{Collection step}\\ 
 buckets $\gB \leftarrow \{\}$\\ 
 \For{$\mA_i \in \gA$}{
    \For{$\vv \in \Rows(s_i\mA_i/\|\mA\|_1)$ } {
            \eIf{\kwexists $B \in \gB$ \kwst $\similarity(\vv,\sum_{\vu\in B} \vu)  \geq 1-\tau$
            }{\nonl \Comment{$\similarity$ is the cosine similarity}\\
                $B \leftarrow B \cup \{\vv\} $\\
                }
            {
            \kwnew $B \leftarrow \{\vv\}$\\
            $\gB \leftarrow \gB \cup \{B\}$\\
            }}
     } 
     $\mA \leftarrow \Mat\left(\{\sum_{\vu \in B}\vu \;|\; B \in \gB \}\right)$ \\\nonl\Comment{$\Mat$ converts a set of vectors into a matrix, each row of $\mA$ is the sum of vectors in a bucket}

\nonl \textbf{Measure step} \\

$\vy \leftarrow \mA\vx + \Lap(\frac{1}{\epsilon}\norm{\mA}_1)$\\

\nonl \textbf{Reconstruct step} \\
$\Bar{\vx} \leftarrow \mA^+ \vy$\\
ans $\leftarrow \{W_i(\Bar{\vx})\;|\; W_i \in \gW\}$\\

\Return ans
 \caption{$\tau$ - Waterfilling Mechanism }
 \label{alg:waterfilling}
\end{algorithm}

 Here we prove a stronger property than either the Sharing Incentive or Non-Interference. We show that adding an additional analyst to an arbitrary collective increase the error experienced by any analyst, A property we call Analyst Monotonicity. 
\begin{theorem} \label{thrm:waterfilling}
Let  $\gW$ be the set of all workloads of the analysts in an arbitrary collective.
 For all analysts $i \neq j$, for all workloads $W_i \in \gW, W_j \not \in  \gW$ the 0-Waterfilling mechanism satisfies both of the following \\ 
\begin{equation} 
    \Err_i\left(\gM, \gW \cup W_j, \left[s_j +\sum_{l: W_l \in \gW}s_l\right] \epsilon\right) \leq \Err_i\left(\gM, \gW , \left[\sum_{l: W_l \in \gW}s_l\right]\epsilon\right)\label{eqn:prf1}
\end{equation}
\begin{equation} \label{eqn:prf2}
    \Err_j\left(\gM, \gW \cup W_j, \left[s_j +\sum_{l: W_l \in \gW}s_l \right] \epsilon\right) \leq \Err_j\left(\gM,W_j, s_j\epsilon\right)
\end{equation}

\end{theorem}
 We first show that regardless of the number of analysts in the collective the scale of the noise added to the queries remains the same. We then show that the error introduced by reconstructing the original query answers (frobenius norm term of \cref{eqn:error}) can only decrease as more analysts are added the collective therefore resulting in error that either decreases or remains the same for each analyst. 

\begin{lemma}
Consider adding an analyst to the collective with strategy matrix $\mA_i$ and weight $s_i$. If the L1 norm of every column of $\mA_i$ is 1, the sensitivity of the resultant strategy queries will increase by $s_i$, formally
\begin{equation*}
     \|\mA'\|_1 =  \|\mA\|_1+s_i,
\end{equation*}
where $\mA$ and $\mA'$ are the resultant strategy matrix before and after adding this analyst respectively.
\label{Lemma:water_sensitivity}
\end{lemma}
\begin{proof}
For any matrix $\mM$, We define $\cnorm(\mM)$ as a vector where the $i$th entry is the L1 norm of the $i$th column of $\mM$, formally
 \begin{equation*}
     \cnorm(\mM) = \sum_{\vv\in\mM} |\vv|,
 \end{equation*}
where $\vv$'s are the row vectors of $\mM$ and $|\vv|$ is the vector which takes entry-wise absolute value of $\vv$.

In Alg.~\ref{alg:waterfilling}, each row of $\mA$ corresponds to a bucket $B \in \gB$. Thus, particularly for $\mA$,
\begin{equation}
    \cnorm(\mA) = \sum_{\vv\in\mA} |\vv| = \sum_{B\in \gB}\left|\sum_{\vu\in B}\vu\right|.
    \label{eqn:water_strategy_bucket}
\end{equation}
Consider adding a query $\vv'$ to buckets $\gB$ and let the new buckets be $\gB'$. Let $\ve' = \vv'/\|\vv'\|$. If $\ve'\cdot \ve_B < 1$ for all buckets $B \in \gB$, $\vv'$ will be put in a new bucket $B'$ and thus $|\sum_{\vu\in B'}\vu| = |\vv'|$. Also, $\gB' = \gB \cup \{B'\}$.

Otherwise, there exists a bucket $B^* \in \gB$ and $\ve' \cdot \ve_{B^*} = 1$. In this case, $\vv'$ will be put in the bucket $B^*$ and $\gB' = \gB$ with updated $B^{*'}$. Since $\ve'$ and $\ve_{B^*}$ are both unit vector, $\ve' \cdot \ve_{B^*} = 1$ means $\vv'/\|\vv'\|=\ve' = \ve_{B^*} = \sum_{\vu \in B^*}\vu/\|\sum_{\vu \in B^*}\vu\|$. Thus,
\begin{equation*}
    \left|\sum_{\vu\in B^{*'}}\vu\right| = \left|\sum_{\vu\in B^*}\vu + \vv'\right| = \left|\sum_{\vu\in B^*}\vu\right| + |\vv'|.
\end{equation*}

In both cases, we have
\begin{equation}
\sum_{B\in \gB'}\left|\sum_{\vu\in B}\vu\right| = \sum_{B\in \gB}\left|\sum_{\vu\in B}\vu\right| + |\vv'|.
\label{eqn:water_add_query}
\end{equation}
In this process, we add $s_i\mA_i$ to $B$ resulting in $B'$. From \cref{eqn:water_strategy_bucket} and \cref{eqn:water_add_query} we get,
\begin{equation*}
\begin{split}
    \cnorm(\mA') &= \sum_{B\in \gB'}\left|\sum_{\vu\in B}\vu\right| = \sum_{B\in \gB}\left|\sum_{\vu\in B}\vu\right| + \sum_{\vv\in s_i\mA_i}|\vv|\\ &= \cnorm(\mA)+\cnorm(s_i\mA_i).
\end{split}
\end{equation*}

Given the L1 norm of every column of $\mA_i \in \gA$ is 1, we have $\cnorm(s_i\mA_i) = s_i\vone$, where $\vone$ is a all-one vector. Since the L1 norm of a matrix is the maximum of all L1 column norms, we have
\begin{equation*}
\begin{split}
    \|\mA'\|_1 &= \max(\cnorm(\mA')) = \max(\cnorm(\mA)+s_i\vone)\\ &= \max(\cnorm(\mA))+s_i = \|\mA\|_1+s_i
\end{split}
\end{equation*}
\end{proof}
Since we can consider the strategy matrix with no analysts as the zero matrix, 
and adding an additional analyst adds their weight to the sensitivity, the L1 norm for the strategy matrix for $k$ analysts is 
\begin{equation*}
    \|\mA\|_1 = \sum_{i=1}^k s_i
\end{equation*}

Since the ith analyst is entitled to $s_i\epsilon$ of the budget and the sensitivity of the strategy query set is equal to the sum of each analysts' weights, the scale of the noise term in \cref{eqn:error} is the same regardless of the number of analysts. Let $z\leq k$ be any arbitrary number of analysts. The scale of the noise term in \cref{eqn:error} is as follows. 
\begin{equation} \label{eqn:noise}
    \frac{2\norm{\bm{A}}_1^2}{\epsilon^2} = \frac{2\left(\sum_{i=1}^z s_i\right)^2}{\left(\sum_{i=1}^z s_i \epsilon\right)^2} = \frac{2}{\epsilon^2}
\end{equation}

Since the amount of noise being added to each query in the final strategy is the same, the amount of error experienced by each analyst is only dependent on the frobenius norm term of \cref{eqn:error}.

We first note that adding a new analyst to the collective results in a change to the overall strategy matrix that can either be expressed by multiplying it by some diagonal matrix with all entries greater than 1 (adding weight to a bucket) or by adding additional rows (creating new buckets). We show below that either of these operations results in a frobenius norm term that is no greater than the term with the original strategy matrix.
\begin{lemma} \label{lemma:diagonal}
For any workload matrix $\mW$ and any strategy $\mA$ \begin{equation*}
    \left\| \mW(\mD\mA)^+\right\|_F \leq \left\| \mW\mA^+\right\|_F
\end{equation*} where $\mD$ is a diagonal matrix with all diagonal entries greater than or equal to 1 and $\mA$ is a full rank matrix.
\end{lemma}
\begin{proof}
We first note that since $\mD$ is a diagonal matrix with all entries greater than or equal to 1 then $\mD^{-1}$ is a diagonal matrix with all values less than or equal to 1. Since this matrix cannot increase the value of any entry of any matrix multiplied by it the following holds. \begin{equation*}
    \left\|\mW\mA^+\mD^{-1} \right\|_F \leq \left\|\mW\mA^+ \right\|_F 
\end{equation*}
We then note that $\mW\mA^+\mD^{-1}$ is a solution to the linear system of equations $\mB(\mD\mA) = \mW$.
Since $\mW\mA^+\mD^+$ is a solution to the linear system of equations then it is the least squares solution to the set of linear equations \cite{pseudo-inverse} and as such the following holds. \begin{equation*}
    \left\|\mW(\mD\mA)^+ \right\|_F \leq \left\|\mW\mA^+\mD^{-1} \right\|_F \leq \left\| \mW\mA^+\right\|_F
\end{equation*}
\end{proof}
\begin{lemma} \label{lemma:rows}
Let $\tilde{\mA}$ be the original strategy matrix $\mA$ with additional queries (rows) added to it. We can write this as a block matrix as  $ \tilde{\mA}  = \begin{bmatrix}
\mA \\
\mC
\end{bmatrix} $ 
Where $\mC$ are the additional queries.
For any workload $\mW$ and any strategy $\mA$
\begin{equation*}
    \left\| \mW\tilde{\mA}^+ \right\|_F \leq \left\| \mW\mA^+ \right\|_F
\end{equation*} 
\end{lemma}
\begin{proof}

Let $\hat{\mA}$ be the original matrix $\mA$ padded with additional rows of zeros in order to be the same size as $\tilde{\mA}$ written in block matrix form as $ \hat{\mA}  = \begin{bmatrix}
\mA \\
\vzero
\end{bmatrix}$.
We note that by the formula for block matrix pseudo-inverse, the pseudo-inverse of $\hat{\mA}$ is as follows. $\hat{\mA}^+  = \begin{bmatrix}
\mA^+ & \vzero
\end{bmatrix}$
We then note that $\mW\hat{\mA}^+$ is a solution to the linear system of equations as follows.

\begin{equation*} \label{eq1}
\begin{split}
\mW\hat{\mA}^+\tilde{\mA} & = \mW \begin{bmatrix}
\mA^+ & \vzero
\end{bmatrix}\begin{bmatrix}
\mA \\
\mC
\end{bmatrix}
 \ =\mW\mA^+\mA 
 \ = \mW
\end{split}
\end{equation*}

Therefore since $\mW\hat{\mA}^+$ is a solution to the linear system of equations and since $ \mW\tilde{\mA}^+ $ is the least squares solution to the linear set of equations \cite{pseudo-inverse} we get the following. 
\begin{equation*}
    \left\| \mW\tilde{\mA}^+ \right\|_F \leq \left\| \mW\hat{\mA}^+ \right\|_F = \left\| \mW\mA^+ \right\|_F
\end{equation*}
\end{proof}

\begin{proof}[Proof of \cref{thrm:waterfilling}]
Let $\mA$ be the strategy matrix produced by the Waterfilling Mechanism without analyst $j$. Let $\tilde{\mA}$ be $\mA$ with additional rows appended to it and let $\mD$ be a diagonal matrix with all entries 1 or greater.
\begin{eqnarray*}
\lefteqn{\Err_i \left(\gM, \gW \cup W_j, \left[s_j+ \sum_{l: W_l \in \gW}s_l \right] \epsilon \right)}  \\
& = & \frac{2}{\epsilon^2}\norm{\mW_i(\mD\tilde{\mA}^+)}_F^2 \ \ \ \text{(from \cref{eqn:noise})} \\
& \leq &\frac{2}{\epsilon^2}\norm{\mW_i\tilde{\mA}^+}_F^2 \ \ \ \text{(from \cref{lemma:diagonal})} \\ 
& \leq& \frac{2}{\epsilon^2}\norm{\mW_i\mA^+}_F^2 \ \ \ \text{(from \cref{lemma:rows})} \\
&  = &\Err_i\left(\gM, \gW, \left[\sum_{l: W_l \in \gW}s_l \right] \epsilon\right) 
\end{eqnarray*}

If we instead assume $\mA$ is the strategy matrix produced by the Waterfilling Mechanism with only analyst $j$ then the same process satisfies \cref{eqn:prf2}.
\end{proof}
Since adding an additional analyst to the collective can only decrease the amount of expected error experienced by any analyst, we have the following as corollaries for Theorem~\ref{thrm:waterfilling}.
\begin{corollary} \label{cor:sharing_incentive}
Waterfilling Mechanism satisfies sharing incentive
\end{corollary}
\begin{corollary}\label{cor:non_interference}
Waterfilling Mechanism satisfies non-interference
\end{corollary}
Unlike Independent Mechanisms, Waterfilling Mechanisms satisfy all the desiderata while being efficient with respect to  error. 
\begin{theorem}
The Waterfilling Mechanism can achieve as much as k times better error than the Independent Mechanism and always achieves no more error than the Independent Mechanism.
\end{theorem}
\begin{proof}
Consider the pathological example of $k$ analysts each of whom ask the same single linear counting query to be answered with the Laplace Mechanism. In this case the overall expected error using the Waterfilling mechanism is that of answering the single query once using the entire privacy budget using the Laplace mechanism. This results in an expected error of $\frac{2}{\epsilon^2}$. If each analyst were to independently answer their queries using $\frac{\epsilon}{k}$ of the budget each and then post process the $k$ results by  taking the sample median it would result in a mean squared error of $\frac{2k}{\epsilon^2}$.
By \cref{cor:sharing_incentive} the Waterfilling Mechanism always achieves at most as much error as the Independent Mechanism satisfying the second statement.
\end{proof}

%% file: experiments.tex
We designed experiments to both test if the mechanisms proposed satisfy the desiderata as well as how they perform in practice. We show 4 different experiments using different inputs and data sets.
\begin{itemize}
\item \textbf{Practical Settings}: We show that the Waterfilling Mechanism maintains high efficiency while still satisfying all three desiderata. We also show that mechanisms that optimize for overall error such as the Utilitarian mechanism fail to satisfy both the Sharing Incentive and Non-Interference.
\item \textbf{Marginals}: Here we show that non-adaptive mechanisms such as the Identity mechanism may incur high error on particular classes of queries such as marginal queries, while adaptive mechanism can perform well on wide ranges of queries.
\item  \textbf{Data-Dependent Non-linear Queries}: We show that the Waterfilling Mechanism retains it's properties when used to reconstruct non-linear queries from a set of linear strategy queries.
\item \textbf{Tolerance for Waterfilling}: We evaluate the efficacy and properties of the mechanism using various levels of $\tau$ and show that $\tau = 10^{-3}$ performs well and does not result in any violations of the sharing incentive.
\end{itemize}

\subsection{Experimental Setup}
\label{sec:Experiments_setup}
 For the following experiments we use HDMM \cite{HDMM} as the selection step, but any selection step can be used in practice. In addition, we can consider the Identity Mechanism a variant of matrix mechanism with a fixed identity strategy matrix $\mI$, $\MM(\mI)$.

For all experiments we used  $\epsilon =1 $ for our total privacy budget. In addition, The Waterfilling Mechanism has a tolerance parameter $\tau$. We experimented with several values of $\tau$. Results shown in \cref{sec:Experiments_Tolerance_results} found $\tau = 0.001$ is a value that achieves good overall accuracy. As such we set it to be $0.001$ in all our experiments.

For the figures, each workload is given an abbreviations as follows: Ind (Independent HDMM), Iden (Identity mechanism), Util (Utilitarian HDMM), WUtil (Weighted Utilitarian HDMM), and Water (HDMM Waterfilling Mechanism). For each experiment we run the optimization 10 times and pick the strategy with the minimum loss. 
\subsection{Empirical Measures}
We design several empirical measures based on our desiderata to provide an overall understanding of the mechanisms. All measures are with respect to a single mechanism and a single set of workloads.

\noindent\textbf{Total Error} is the sum of expected errors of all analysts. This is a common measure found in the literature to show the efficiency of the algorithm. \par 

\noindent\textbf{Maximum Ratio Error} of a mechanism $\gM$ for a given analyst is the expected error of $\gM$ divided by the expected error of the independent version. For non-independent adaptive algorithms, it is a measure of the Sharing Incentive as it measures to what extent one analyst gets better or worse off compared to asking the query on their own. We present the maximum of the ratio errors among all analysts. The maximum ratio error amongst all analysts is
\begin{equation*}
    \max_i \left(  \frac{\Err_i(\gM,\gW, \epsilon)}{\Err_i(\gM,W_i, s_i\epsilon)} \right).
\end{equation*} 
If the value is larger than 1, the mechanism violates the Sharing Incentive as the error in the joint case is greater than the error experienced in the independent case. 

\par
\noindent\textbf{Empirical Interference} is a quantifiable measure to show the extent which a mechanism violates Non-Interference or the distance from violating it. For each analyst $i$, we define the interference with respect to another analysts $j$ as the ratio of the expected error for analyst $j$ when all analysts are included to the case when excluding analyst $i$. If this ratio is larger than 1, analyst $j$ can be worse off when analyst $i$ joins the workload set.  We define the interference of analyst $i$ on analyst $j$ to be
\begin{equation*}
    I_{i}(j) = \frac{\Err_j(\gM,\gW, \epsilon)}{\Err_j(\gM,\gW^c_i, (1-s_i)\epsilon)}
\end{equation*} 
This represents the relative change in error experienced by analyst $j$ when analyst $i$ joins the collective. We then define the interference of mechanism $\gM$ on the set $\gW$ as the maximum of interference among all analysts, as
\begin{equation*}
    I_{\gM}(\gW) = \max_{1 \leq i,j \leq k, i\neq j} I_{i}(j).
\end{equation*} 

Intuitively, it represents the maximum ratio increase of the expected error of any analyst when another analyst joins the workload set. If $I_{\gM}(\gW) \leq 1$, mechanism $\gM$ satisfies Non-Interference on $\gW$.
Since $\gM$ is usually a non-deterministic mechanism, rerunning the mechanism with $\gW^c_i$ may give different strategy matrices to other analysts. Thus, we fix strategy matrices for Select First Mechanisms to ensure a more reasonable comparison. Since the strategies used by Collect First Mechanisms are dependent on each analysts input it is not possible to fix the strategy matrix.

\subsection{Workloads and Datasets}
\label{sec:Experiments_WorkloadsAndDatasets}
Here we describe the methods used to generate workloads for each analyst as well as the data-sets used. When considering only linear queries all of our mechanisms are data independent and as such do not require a dataset in order to be evaluated. We only use a dataset when we extend our evaluation to non-linear queries and data dependent queries.  \par

\noindent\textbf{Practical settings:}\label{sec:prac_setup}\input{experiment_figure} 
We generate practical settings using a series of random steps using the census example workloads provided in \cite{HDMM}. We tested on the race workloads with domain size $n=64$.
\begin{enumerate}
    \item We first fix the domain size $n$. We then generate the number of analysts by picking an integer $k$ uniformly random from $[2, k_{\max}]$. We let the number of analysts be $k$. Each analyst is given equal weight.
    \item Each analyst then pick a workload uniformly random from the set of 8 workloads, including 3 race workloads, Identity, Total, Prefix Sum, H2 workload, and custom workload.
    \item If they get custom workload, we chose their matrix size by picking an integer uniformly random from $[1, 2n]$.
    \item For each query in the matrix we chose a class of query uniformly sampled from the set including range queries (0-1 vector with contiguous entries), singleton queries, sum queries (random 0-1 vector) and random queries (random vector). The query is thus a random query within its class.
    \item The custom workload is thus a vertical stack of the queries.
    \item We repeat this procedure $t$ times to get $t$ randomly chosen sets of workloads. We call them $t$ instances.
\end{enumerate}

\noindent\textbf{Marginals:}\label{sec:Experiments_Marginal_setup}
We also experiment on another common type of workloads, marginals. For a dataset with $d$ attributes with domain size $n_i$ for the $i$th attribute, we can define a $m$-way marginal as the follows. Let $S$ be a size $m$ subset of $\{1,2,\ldots,d\}$, we can express the workload as the Kronecker product $\mA_1 \otimes \mA_2 \otimes \ldots \otimes \mA_d$, where $\mA_i = \mI_{n_i}$ if $i \in S$ and $\mA_i = \mT_{n_i}$ otherwise. Here $\mI_{n_i}$ is the identity workload matrix and $\mT_{n_i}$ is the total workload matrix. Specifically, a 0-way marginal is the Total workload and a $d$-way marginal is the Identity workload. Also, since there are $\binom{d}{m}$ size-$m$ subset of $\{1,2,\ldots,d\}$, there are $\binom{d}{m}$ different $m$-way marginals. In our experiments for simplicity, we use $d$ attributes all with domain size 2. We repeat the process for generating analyst workloads from the practical settings in this case each individual analyst chooses a workload uniformly at random from the set of set of $\binom{d}{m}$ $m$-way marginals.

\noindent\textbf{Data-dependent Non-linear Queries: }\label{sec:experiments_nonlinear_setup}In previous experiments, all workloads are linear and the expected error can thus be calculated without data. Our mechanisms can also be used for non-linear queries. We experiment on some common non-linear queries including \emph{mean}, \emph{medium}, and \emph{percentiles} based on a histogram.

Error in this case is data-dependent and needs to be empirically calculated using real datasets. We use the Census Population Projections \cite{census_population}. The dataset is Population Changes by Race. We choose year 2020 and Projected Migration for Two or more races. The  domain size of data is $n=86$, representing ages from 0 to 85.

As in the previous 2 experiments we use the procedure from practical settings in order to generate each analyst's workloads except the set of workloads to select from only contains 4 queries, \emph{mean}, \emph{medium}, \emph{25-percentile}, and \emph{75-percentile}. \emph{Mean} is reconstructed from the workload containing the Total query $\mT_n$ and the weighted sum query, a vector representing the attribute values (0 to 85 in our case). \emph{Medium} and \emph{percentiles} are reconstructed from the Prefix Sum workload $\mP_n$.

\noindent\textbf{Tolerance for Water-filling: }
\label{Sec:Experiments_Tolerance_setup}
To examine the effect of tolerance in practice, we experimented on different values of tolerance $\tau$ for the HDMM Water-Filling mechanism. \cref{fig:tol} shows the case when $\tau \in [0.1, 0]$. We experimented with greater value of $\tau$ those values resulted in greater error and have been omitted from the figures.The workloads used are 1-way marginals as defined in \cref{sec:Experiments_Marginal_setup}.
\subsection{Results}
\noindent\textbf{Practical settings:}
\label{sec:prac_results} \cref{fig:prac_total} gives an overall view of the efficiency of different mechanisms. As expected, Utilitarian HDMM, a mechanism optimized for overall error, performs the best. Meanwhile Independent HDMM, a mechanism which does not utilize the group structure of the problem at all performs the worst. We note that the Weighted Utilitarian Mechanism in exchange for satisfying the sharing incentive performs slightly worse than the Utilitarian but performs better than the Waterfilling Mechanism which satisfies all three desiderata. The Waterfilling Mechanism performs as well as the Identity Mechanism while still satisfying adaptivity. This shows as stated in \cref{sec: Tradeoffs} that while there is a small cost in order to satisfy the sharing incentive and Non-Interference, satisfying adaptivity comes at no accuracy cost.

We present the results for $k_{\max}=20$ as a representative in \cref{fig:prac}. The figure is a box plot of $t=100$ instances is generated randomly using the procedure in \cref{sec:prac_setup}. The green line represents the median and the green triangle represents the mean. The box represents the interquartile range.

\cref{fig:prac_ratio} shows how other mechanisms compared with Independent HDMM in terms of maximum ratio error. Utilitarian HDMM violates the Sharing Incentive in a small number of instances as there are some outliers with maximum ratio error larger than 1. Weighted Utilitarian and The Waterfilling Mechanism satisfied the Sharing Incentive. Although Identity also has some outliers larger than 1, since independent HDMM is not the independent form of this mechanism it does not violate the Sharing Incentive.

\cref{fig:prac_inter} gives an empirical indication on whether a mechanism satisfies Non-Interference. It can be seen that both Utilitarian and Weighted Utilitarian HDMM violate Non-Interference in some cases. Weighted Utilitarian has fewer instances which violate Non-Interference than Utilitarian. The Weighted Utilitarian mechanism also violates Non-Interference to a smaller extent than the Utilitarian Mechanism. The other three mechanisms do not violate Non-Interference as we expect. \par

\noindent\textbf{Marginal Workloads: }
\label{sec:Experiments_Marginal_results}
In \cref{fig:prac} we show the results for $1$-way marginal with $d=8$, $k_{\max}=20$, and $n =256$. This figure also contains 100 instances. In particular, there are $d$ 1-way marginals each corresponds to an attribute.
\cref{fig:marginal_total} shows Identity mechanism performs worse than the Waterfilling Mechanism and both Utilitarian mechanisms. The addition of the 1-way marginals drastically increases the error of identity compared to that of the other mechanisms. This is an example where the Identity Mechanism performs poorly with regard to total error for a common type of workloads. This is also observed for 1-way marginals with $d=6,7,9,10$. 
\cref{fig:marginal_ratio} and \cref{fig:marginal_inter} are qualitatively similar to those in the practical settings. The Waterfilling Mechanism continues to satisfy all the desiderata while maintaining lower error than the Independent and Identity Mechanisms. Both Utilitarian mechanisms achieve lower overall error but at the cost of violating non interference.

\noindent\textbf{Data-dependent Non-linear Queries: } \label{sec:experiments_nonlinear_results}
\begin{figure*}[ht]
\resizebox{0.9\textwidth}{!}{%
	\centering
	\begin{subfigure}[b]{0.308\linewidth}
		\includegraphics[width=1\textwidth]{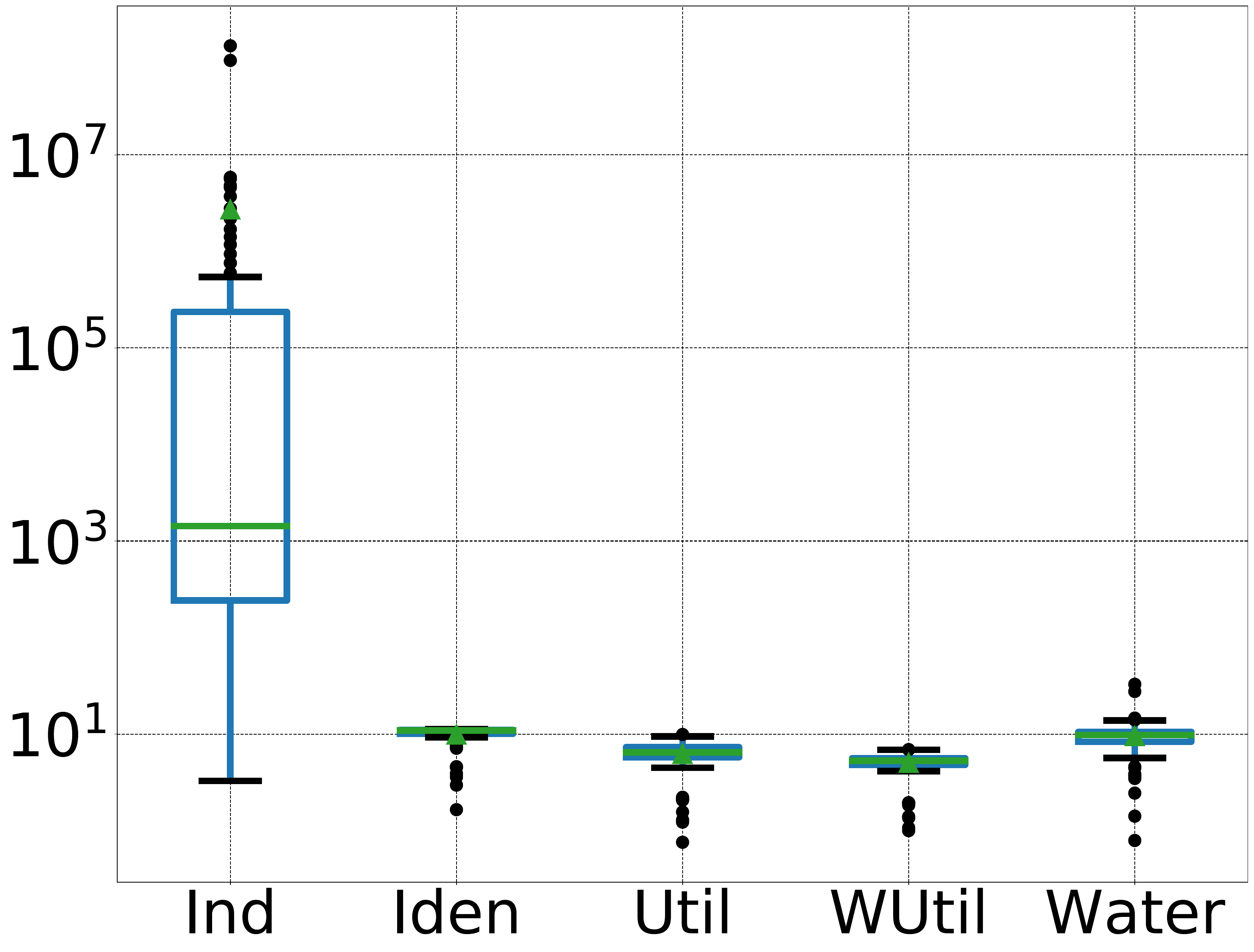}
		\caption{Total Errors (log scale)}
		\label{fig:data_total}
	\end{subfigure}
	\begin{subfigure}[b]{0.254\linewidth}
		\includegraphics[width=1\textwidth]{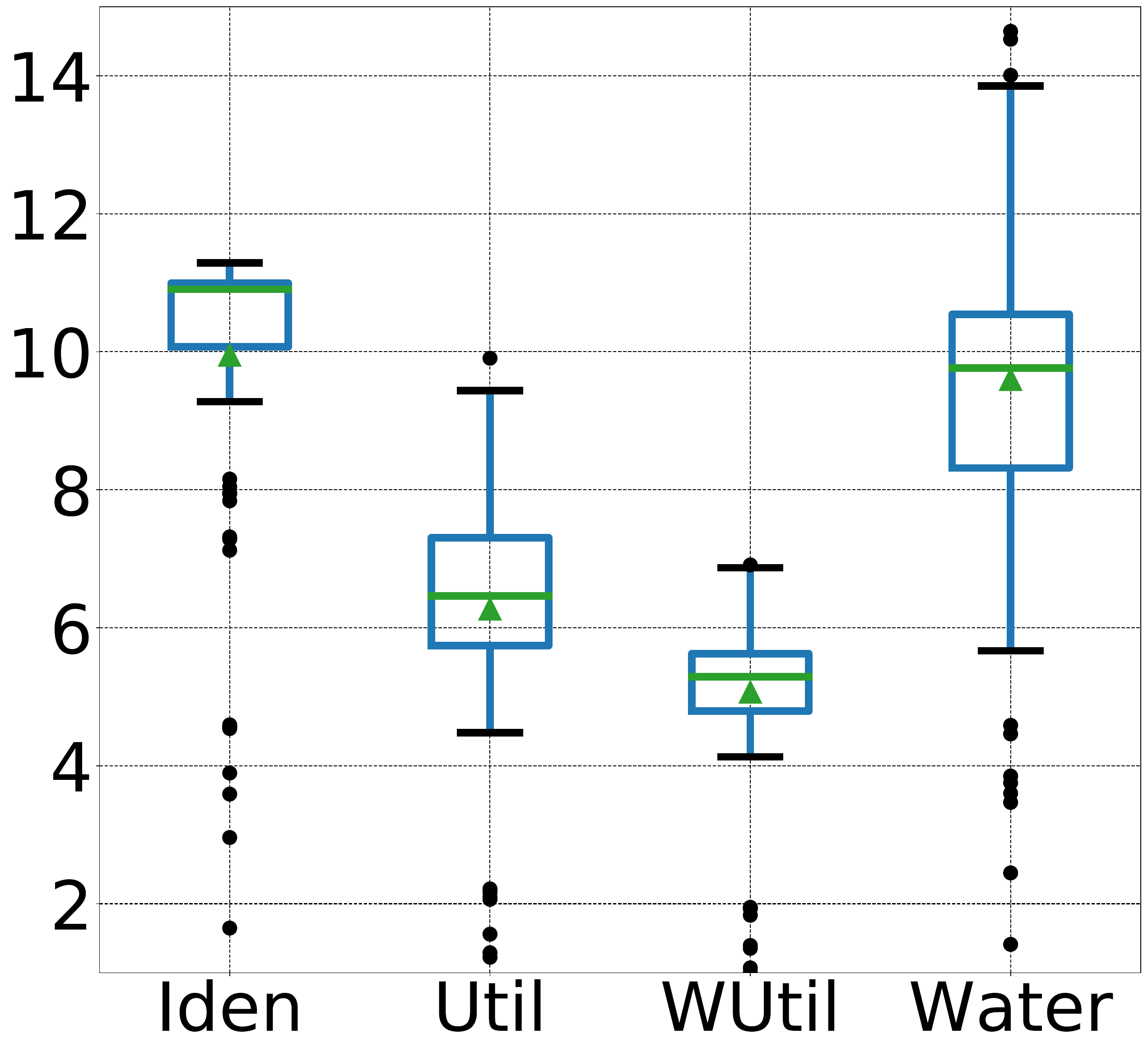}
		\caption{Total Errors (zoomed in)}
		\label{fig:data_total_zoom}
	\end{subfigure}
	\begin{subfigure}[b]{0.217\linewidth}
		\includegraphics[width=1\textwidth]{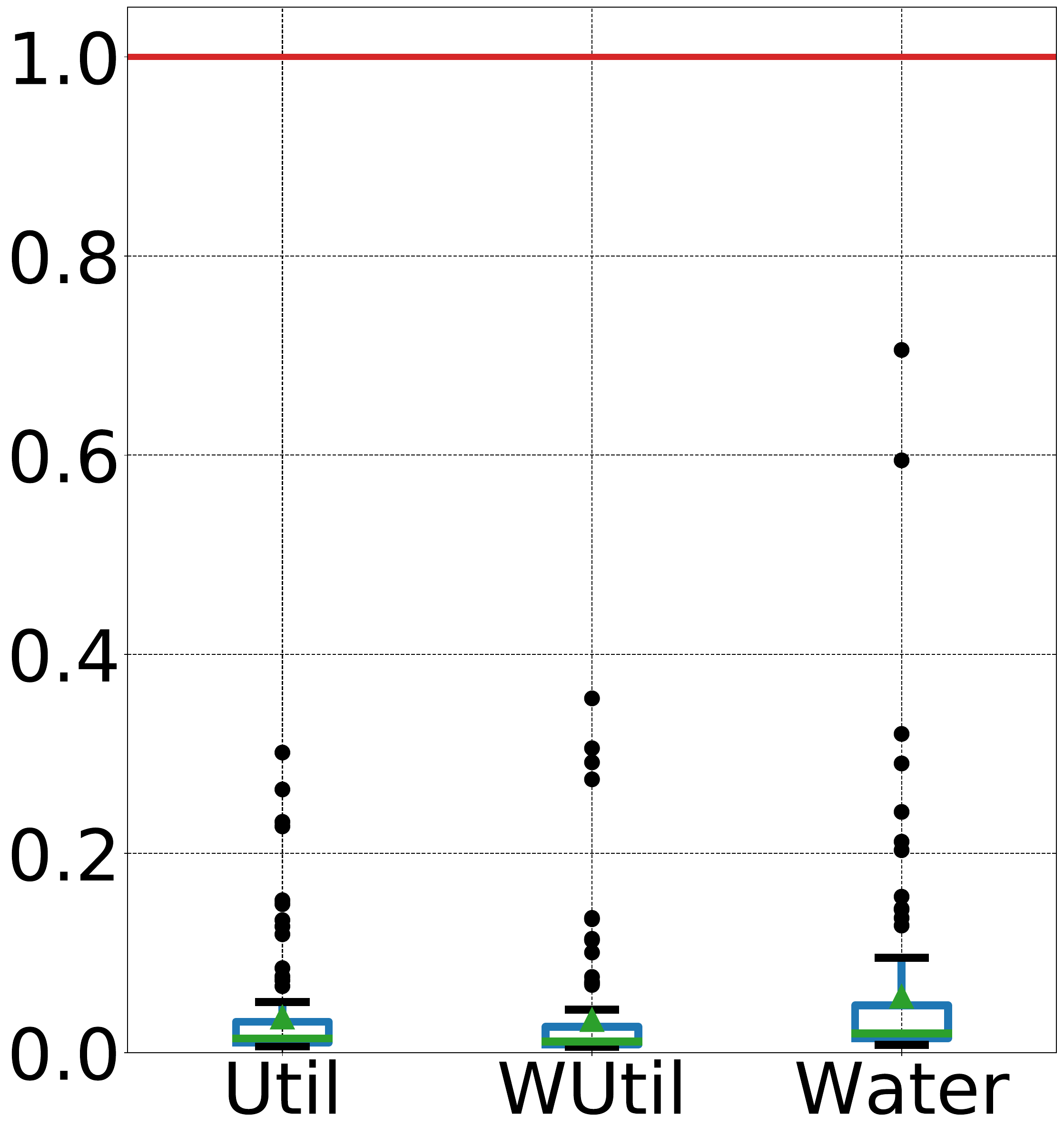}
		\caption{Max Ratio Errors}
		\label{fig:data_ratio}
	\end{subfigure}
	\begin{subfigure}[b]{0.223\linewidth}
		\includegraphics[width=1\textwidth]{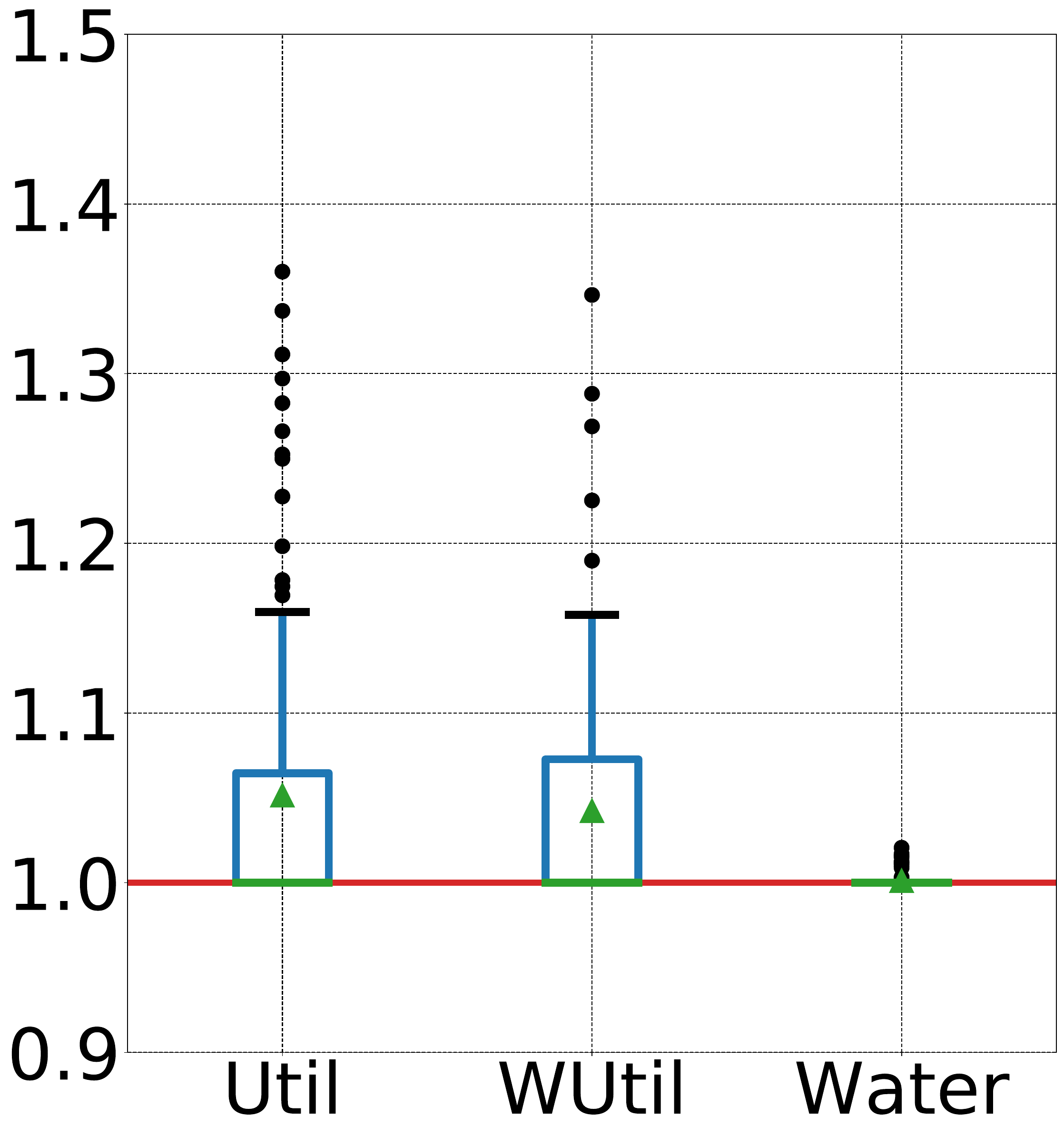}
		\caption{Empirical Interference}
        \label{fig:data_inter}
    \end{subfigure}%
}
\caption{Empirical measures for non-linear queries. Errors shown are empirical expected errors calculated using real data. Values of maximum ratio error and empirical interference above 1 signify a violation of the Sharing Incentive and Non-Interference respectively.}
\label{fig:data}
\end{figure*}
 \cref{fig:data_total} shows that the Independent Mechanism performs much worse than all other mechanisms in terms of total error. \cref{fig:data_total_zoom} is the zoomed in version of \cref{fig:data_total}, removing Independent. Since the answer of a non-linear query is reconstructed using the result of a different linear workload, Utilitarian is not guaranteed to have the lowest total errors. We can see that Weighted Utilitarian outperforms Utilitarian here. The other two mechanism have higher total errors, and the Waterfilling Mechanism has a better median total errors than Identity.

\cref{fig:data_ratio} and \cref{fig:data_inter} shows the max ratio errors and empirical interference. Since Independent and Identity mechanism satisfy the Sharing Incentive and Non-Interference by definition, we omit them here. We can see that all 3 other mechanisms satisfy the Sharing Incentive as they all have max ratio errors smaller than 1. Both Utilitarian mechanisms violate Non-Interference as shown in \cref{fig:data_inter}. Waterfilling Mechanisms satisfies Non-Interference. The outliers are due to numerical errors since we are using empirical expected errors instead of analytical ones.

These results show that our mechanisms also perform well for non-linear queries and have similar properties as the instances with linear queries. The results are qualitatively similar for $k_{\max}=10$.

\begin{figure}[h]
  \begin{subfigure}[b]{0.45\linewidth}
    \centering
    \includegraphics[width=1\linewidth]{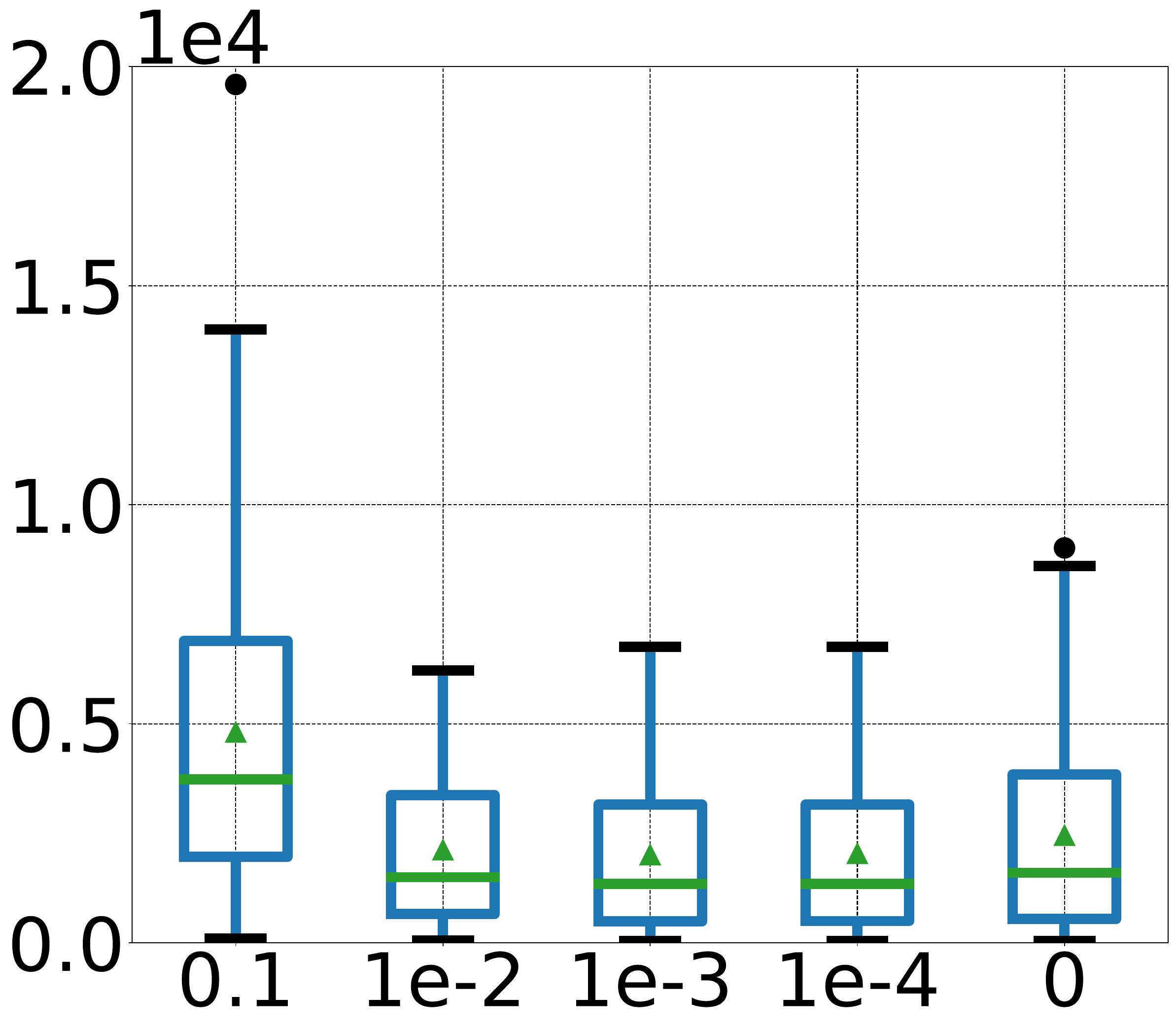}
    \caption{Total Errors}
    \label{fig:tol_total}
  \end{subfigure}
  \begin{subfigure}[b]{0.44\linewidth}
    \centering
    \includegraphics[width=1\linewidth]{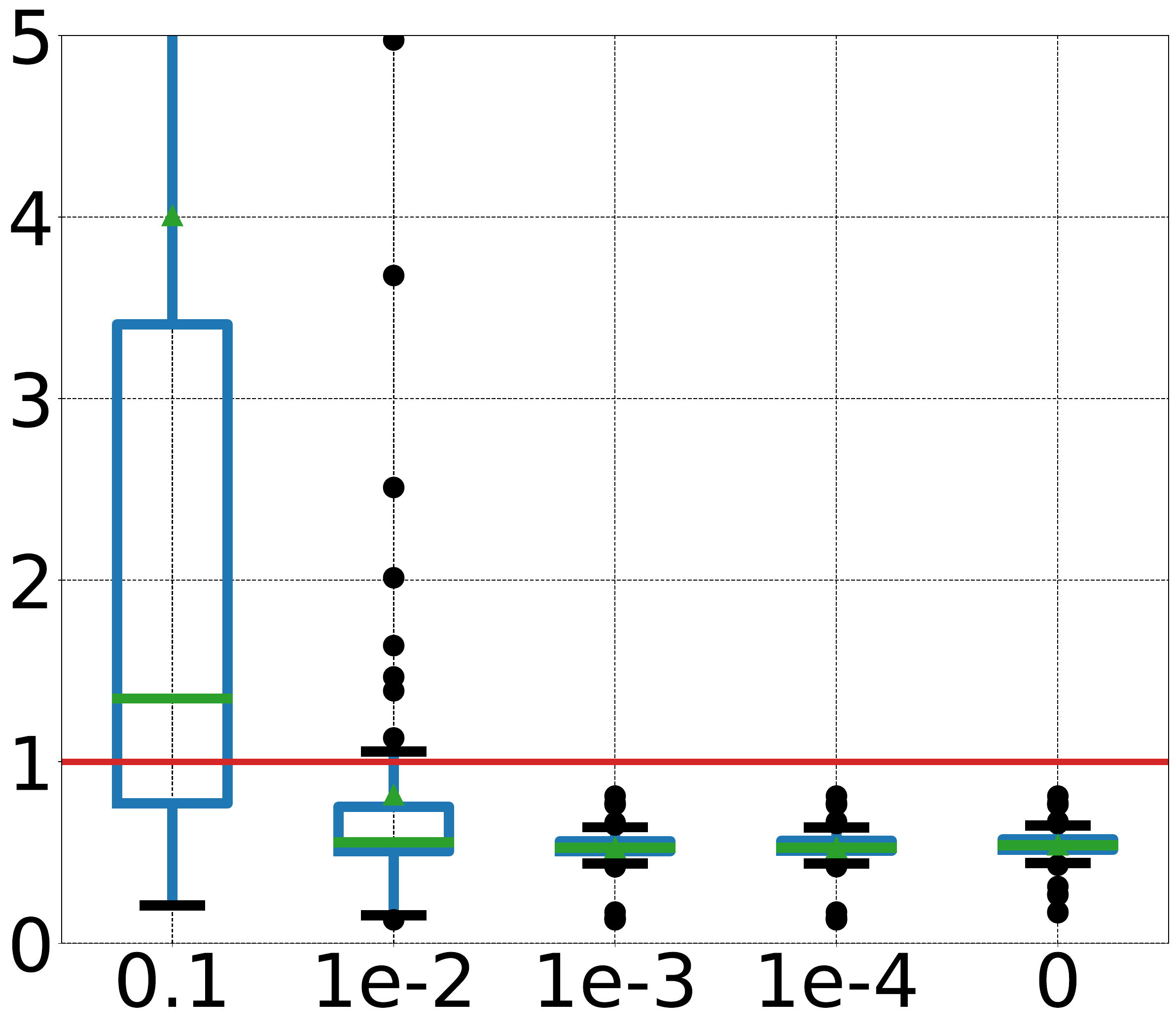}
    \caption{Max Ratio Errors}
    \label{fig:tol_ratio}
  \end{subfigure}
  \caption{Total and Maximum Ratio Errors for 1-way marginals using HDMM Water-filling mechanism with different values of $\tau$ (x-axis). Values of maximum ratio error above 1 signify a violation of Sharing Incentive.}
  \label{fig:tol}
\end{figure}

\noindent\textbf{Tolerance for Water-filling Mechanism: }
\label{sec:Experiments_Tolerance_results}
\cref{fig:tol_ratio} shows that the total error is large at both ends, $\tau =0.1$ and $\tau =0$. The total error is the smallest for $\tau=0.01$ and is also small for $\tau = 10^{-3}$ and $\tau=10^{-4}$. This shows that there is no simple relation between the value of tolerance and total errors and we should not set $\tau=0$ exactly in practice. \cref{fig:tol_ratio} shows the violation of Sharing Incentive when $\tau =0.1$ and $\tau=0.01$. From this result, we see that $\tau=0.01$ is too large and $\tau=10^{-3}$ (our default setting) is reasonable. We do not observe violation of Non-Interference any value of $\tau$.

%% file: experiment_figure.tex
\begin{figure*}[ht]
\resizebox{0.80\textwidth}{!}{%
	\centering
	\begin{subfigure}[b]{0.33\textwidth}
		\includegraphics[width=1\textwidth]{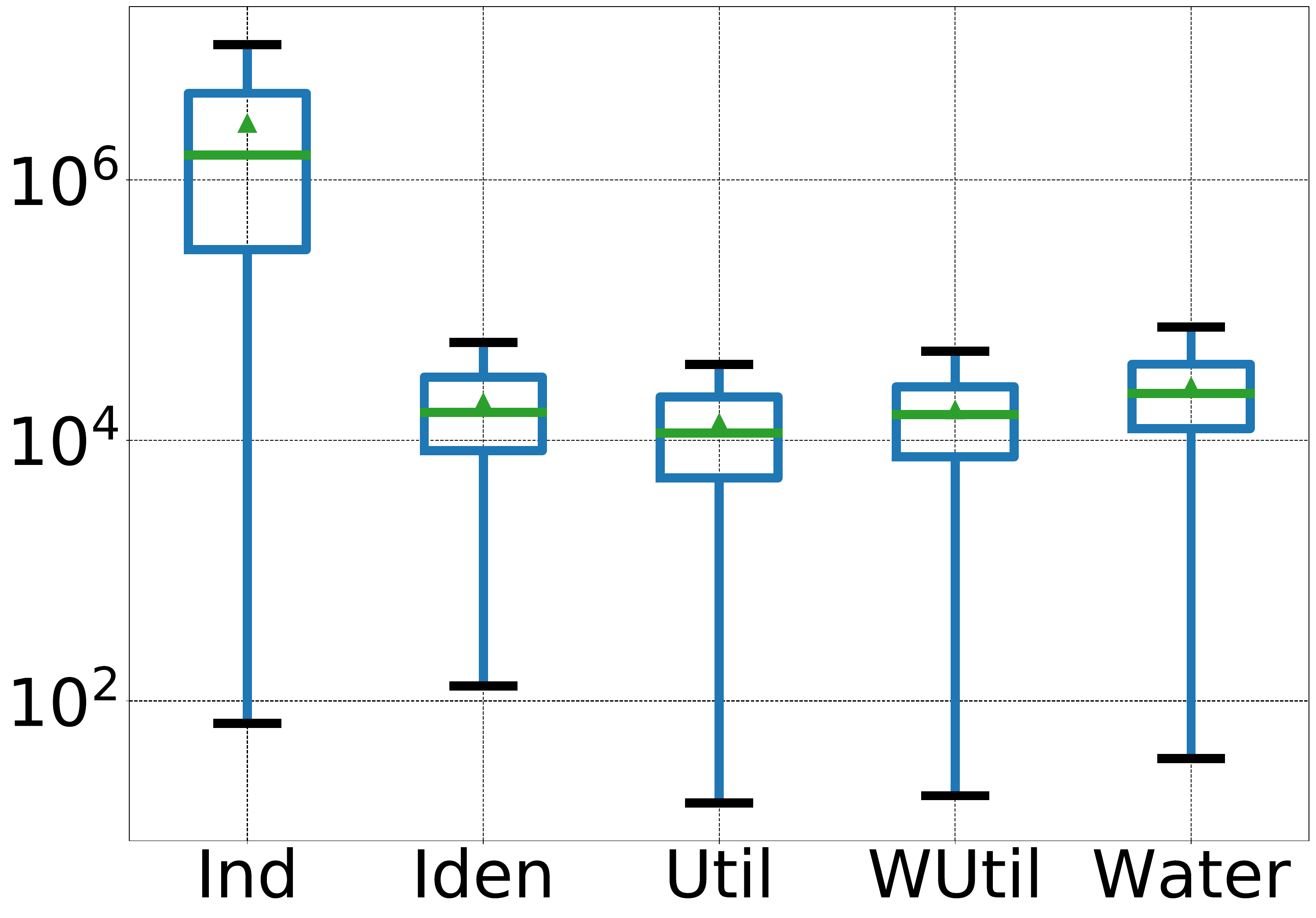}
		\caption{Practical Total Errors (log scale)}
		\label{fig:prac_total}
	\end{subfigure}
	\begin{subfigure}[b]{0.335\textwidth}
		\includegraphics[width=1\textwidth]{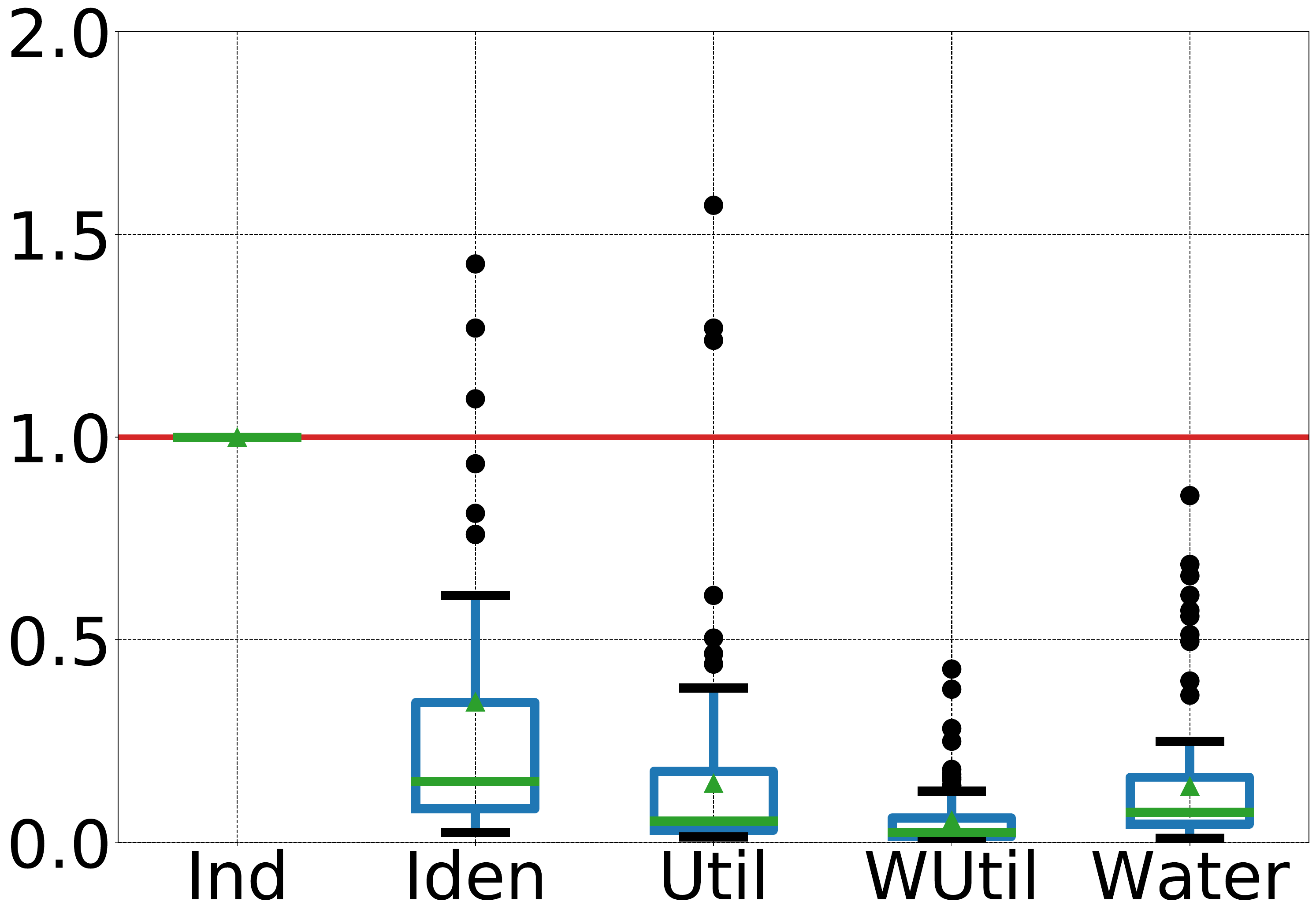}
		\caption{Practical Max Ratio Errors}
		\label{fig:prac_ratio}
	\end{subfigure}
	\begin{subfigure}[b]{0.325\textwidth}
		\includegraphics[width=1\textwidth]{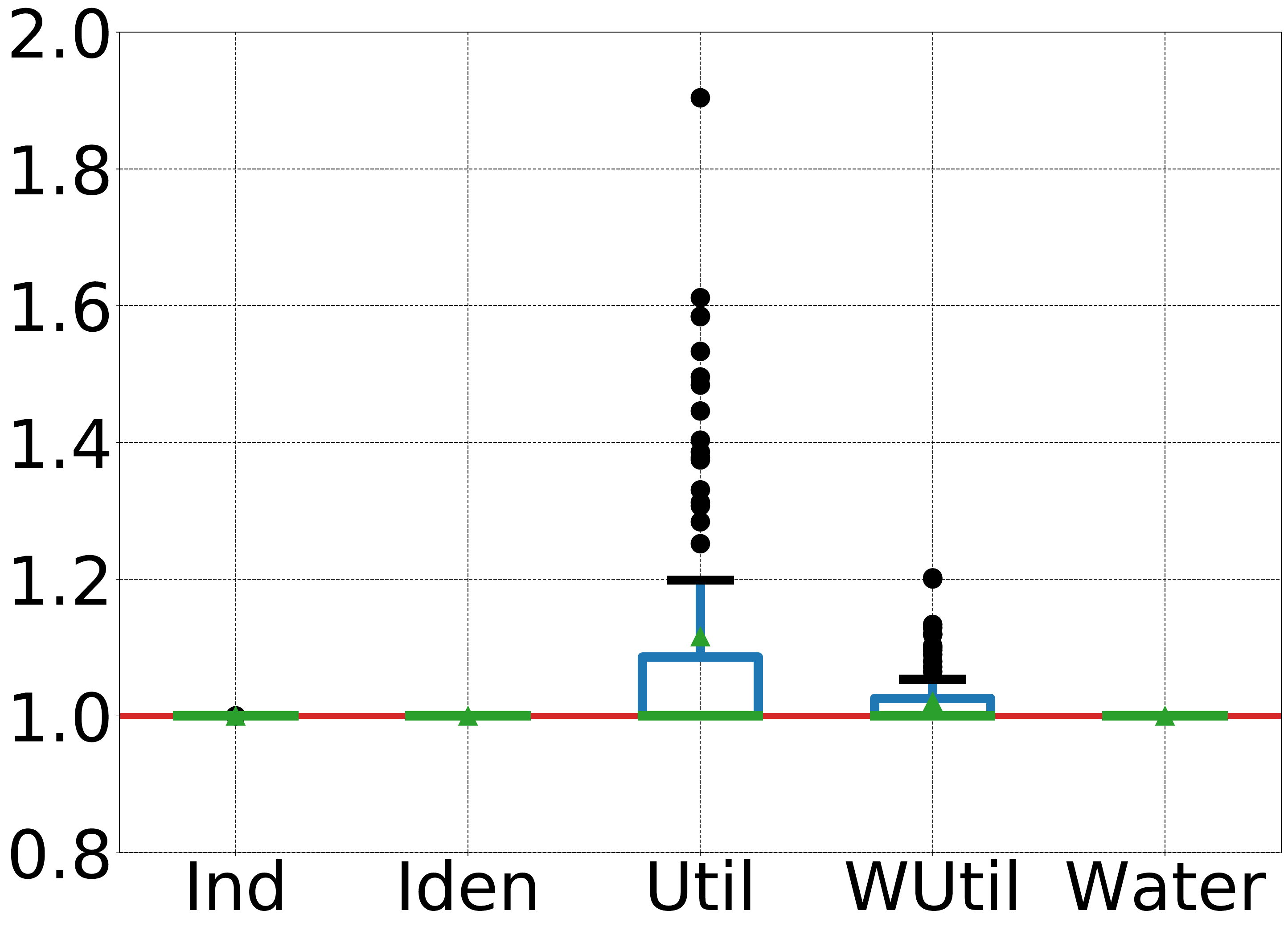}
		\caption{Practical Empirical Interference}
        \label{fig:prac_inter}
    \end{subfigure}%
}

\resizebox{0.80\textwidth}{!}{%
    \begin{subfigure}[b]{0.33\textwidth}
		\includegraphics[width=1\textwidth]{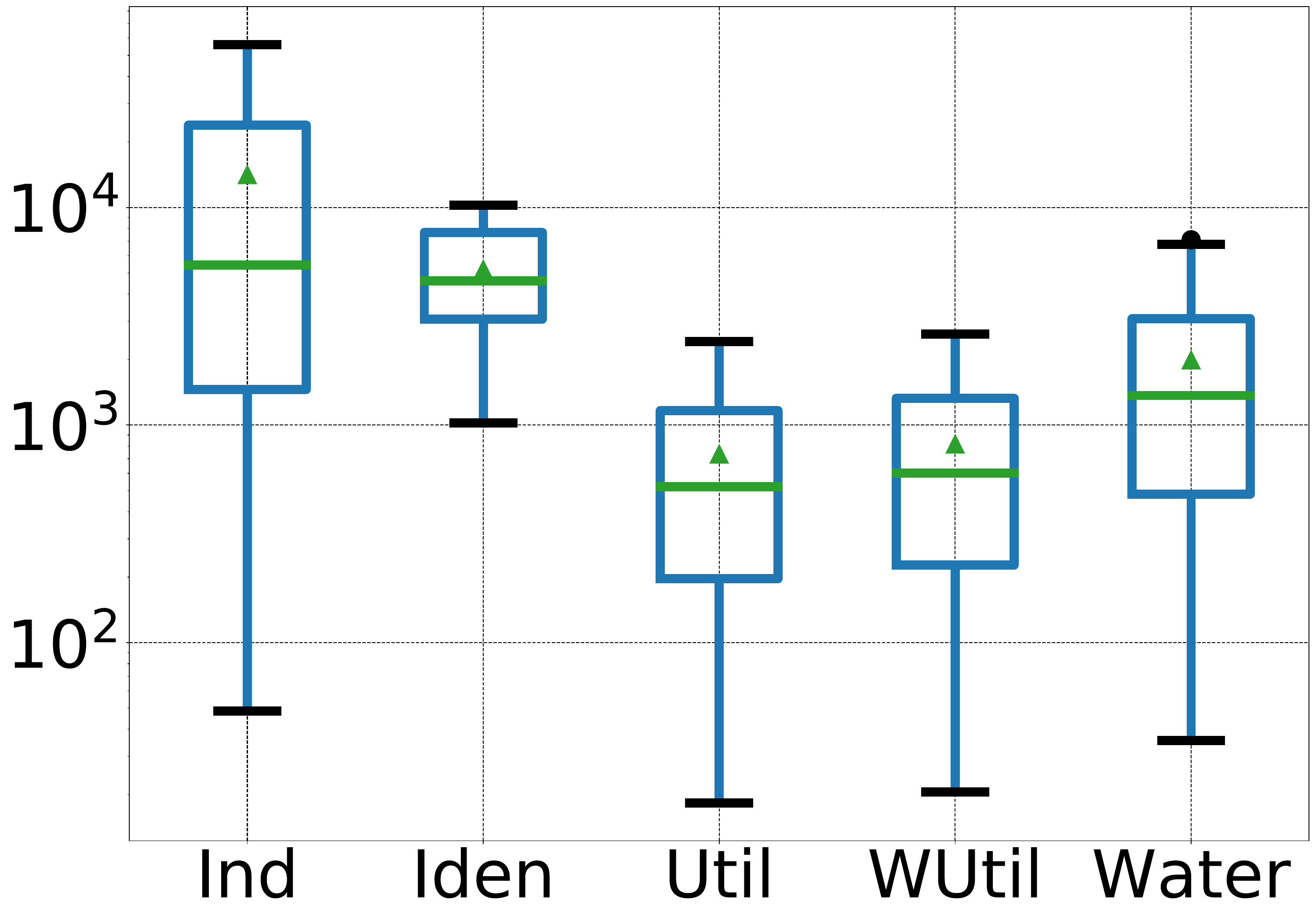}
		\caption{Marginal Total Errors (log scale)}
		\label{fig:marginal_total}
	\end{subfigure}
	\begin{subfigure}[b]{0.335\textwidth}
		\includegraphics[width=1\textwidth]{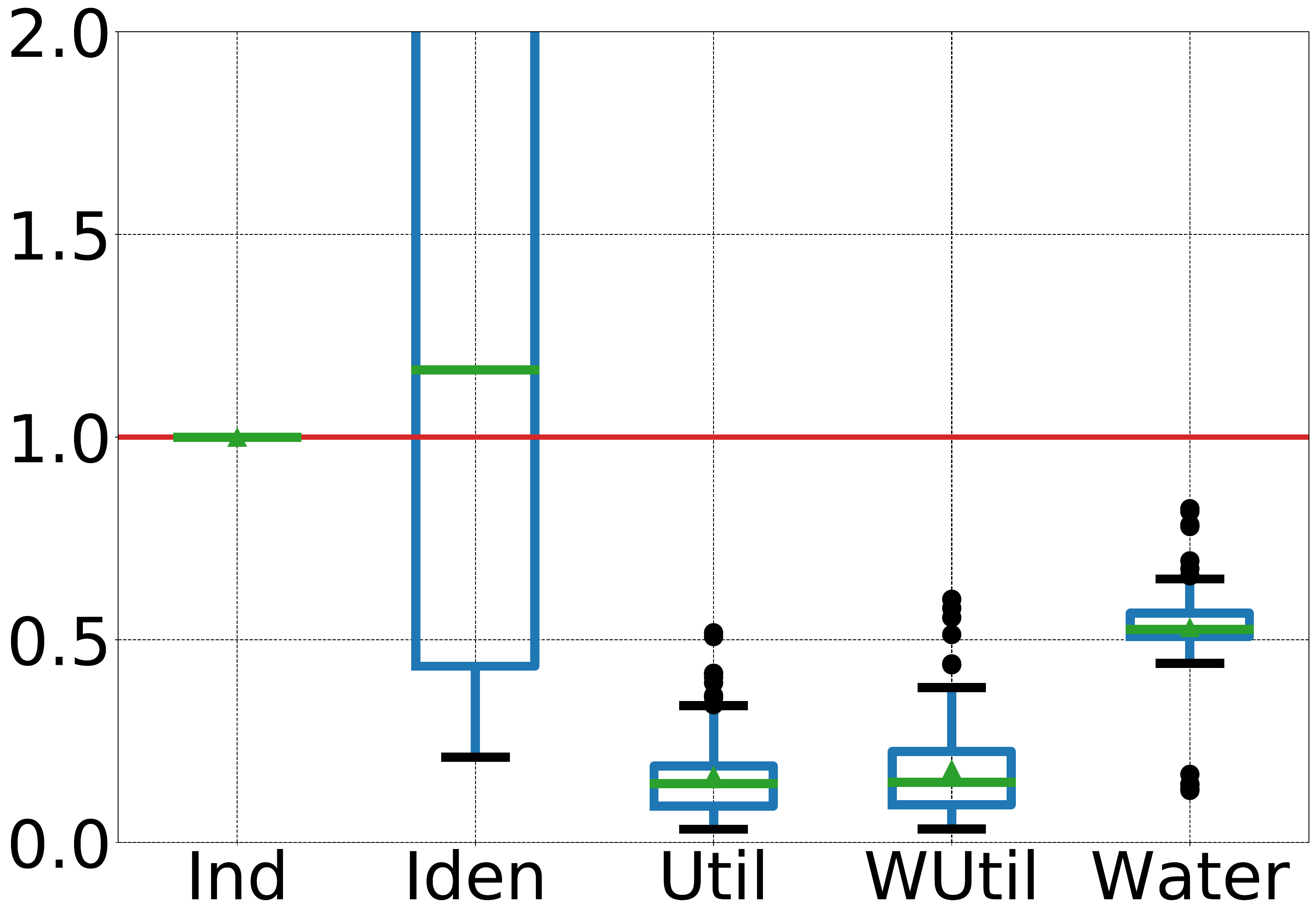}
		\caption{Marginal Max Ratio Errors}
		\label{fig:marginal_ratio}
	\end{subfigure}
	\begin{subfigure}[b]{0.325\textwidth}
		\includegraphics[width=1\textwidth]{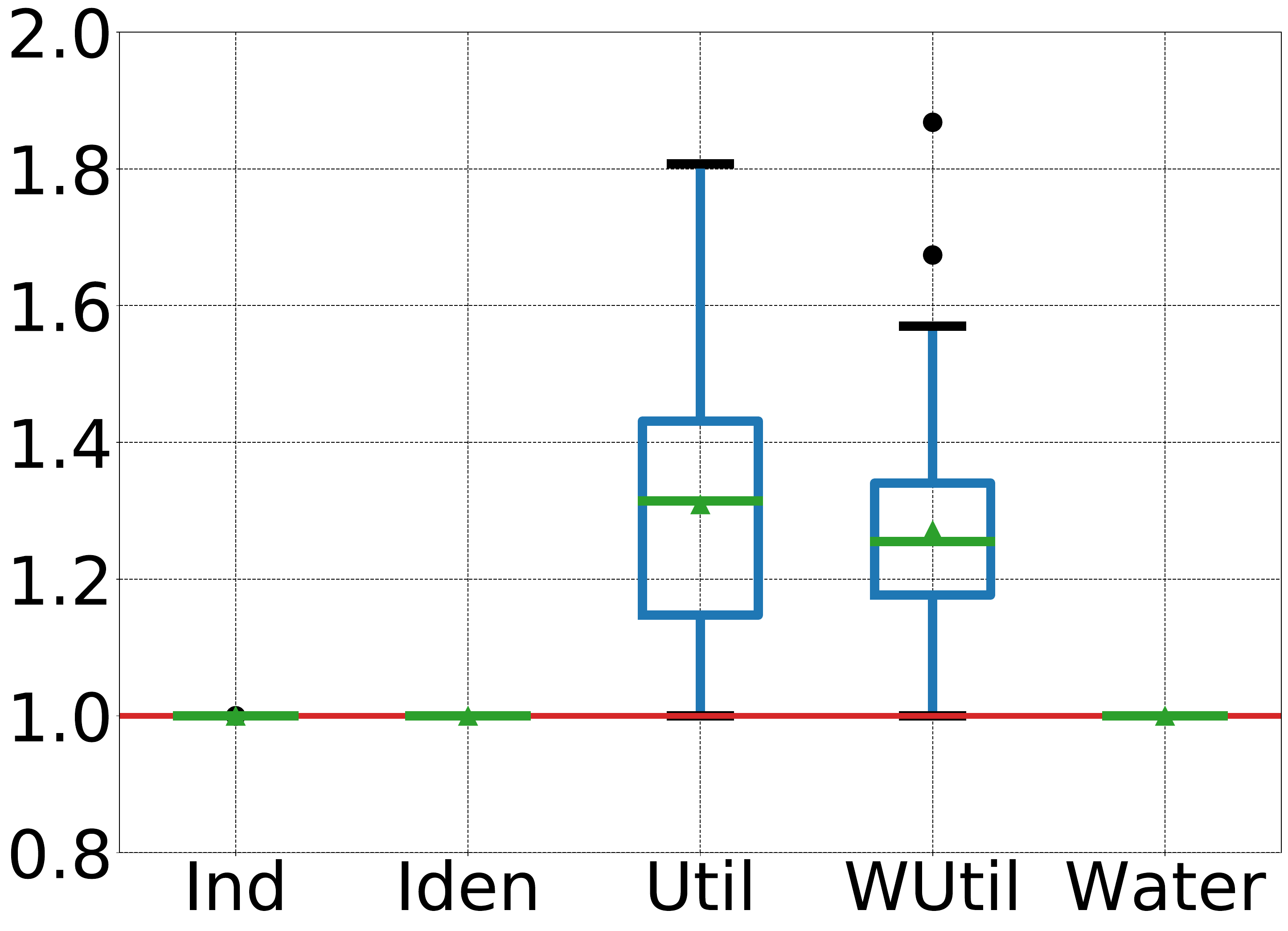}
		\caption{Marginal Empirical Interference}
        \label{fig:marginal_inter}
    \end{subfigure}%
}
\caption{Empirical Measures for practical settings (above) and 1-way marginals (below). Values of maximum ratio error and empirical interference above 1 signify a violation of the Sharing Incentive and Non-Interference respectively.}
\label{fig:prac}
\end{figure*}

%% file: relatedwork.tex
There has already been significant work on answering sets of queries in a differentially private manner, including theoretical lower bounds on error \cite{bhaskara_dadush_krishnaswamy_talwar_2012,hardt_talwar_2010} and many  practical algorithms \cite{Xu2013,Matrix10,HDMM,AHP,Chen13:recursive,Ding2011,Hay2010,Li2014,Narayan,Hb,Acs2012}. Each of these mechanisms primarily attempts to optimize the total error (or utilitarian social welfare) instead of distributing error in some manner. Likewise these mechanisms are not intended for any group answering setting but are instead designed for single analyst use.  \par 
Sharing computational resources such as memory and network has been considered in the context of resource allocation for data centers and networking~\cite{drf, fairCloud, tradeoffs, beyondDRF, dynamicPropSharing, ROBUS, Mesos}. For example, the influential work on dominant resource fairness~\cite{drf} studies the allocation of several heterogeneous computational resources among agents (the owners of various jobs in a data center) and designs protocols that are simultaneously efficient and ensure good treatment of all agents through the Sharing Incentive and strategy-proof guarantees. In a sense, our work considers the same questions of how to design an effective shared system from the perspective of differential privacy and data release, recognizing that in the common case where there are multiple analysts, privacy budget is indeed a shared resource.

%% file: futurework.tex
\label{sec:Future_work}

There remain many technical problems in Differential Privacy which remain unanswered and may serve as powerful tools in the multi analyst setting.
In this work we consider the offline setting where analysts submit their entire workload in advance and receive all of their answers at once. However most query answering settings are done in an online setting where analysts may adaptively chose their next query in response to a previous query answer. While there is some work on online differentially private query answering \cite{PrivateSQL,Flex} there are  still significant hurdles to be overcome.To the best of our knowledge there is no differentially private mechanism which which answers queries with an adaptive strategy optimized to account for arbitrary prior knowledge.  Such a mechanism would be essential to the online multi analyst problem as it would allow for prior query answers to inform future query answers and budget use.

%% file: conclusions.tex
We see as in \cref{fig:prac_total} that the traditional method of independently answering using fractional budgets results in an enormous increase in overall error when compared to joint mechanisms. In our practical cases we see over an order of magnitude difference between independent HDMM and HDMM waterfilling. We show in \cref{fig:prac_ratio} that a naively implemented joint mechanism (utilitarian HDMM) can result in violation of the Sharing Incentive resulting in some analysts gaining their extra utility at the expense of other analysts who are worse off than in the independent case. Likewise \cref{fig:prac_inter} shows that naively implemented joint mechanisms can allow analysts to interfere with other analysts by asking vastly different query sets. In \cref{fig:marginal_total} we show that mechanisms which are non-adaptive may suffer great losses in utility based off the queries being asked. When compared to the Utilitarian mechanism, which directly optimizes on overall error, the Waterfilling mechanism performs slightly worse while still satisfying all the desiderata.

%% file: acknowledgement.tex
This work was supported by DARPA and SPAWAR under contract N66001-15-C-4067.